\documentclass[fleqn,usenatbib,pdfencoding=auto,psdextra]{mnras}

\usepackage[T1]{fontenc}
\usepackage{graphicx}
\usepackage[acronym]{glossaries}
\usepackage{booktabs}
\usepackage{xcolor}
\usepackage{url}
   \usepackage{subcaption}
\usepackage{amsmath}
\usepackage{amssymb}

\hypersetup{
    colorlinks=true,
    linkcolor=blue,
    filecolor=magenta,
    urlcolor=blue,
    citecolor=blue
}

\newacronym{bh}{BH}{}

\newacronym{emris}{EMRIs}{}
\newacronym{emri}{EMRI}{}
\newacronym{mbh}{MBH}{}
\newacronym{gw}{GW}{}
\newacronym{few}{\texttt{few}}{}
\newacronym{psd}{PSD}{}
\newacronym{snr}{SNR}{}
\newacronym{fm}{FM}{}

\newcommand{\ff}[1]{{\color{red}{[FF: #1]}}}
\newcommand{\frd}[1]{{\color{blue}{[Francisco: #1]}}}

\begin{document}

\title[Gotta light? Illuminating AGN disks with LISA EMRIs]
{\textit{Gotta light?} Illuminating AGN disks with LISA EMRIs}

\author[F. Fantoccoli et al.]{
Federico Fantoccoli,$^{1}$\thanks{E-mail: federico.fantoccoli@aei.mpg.de}
Francisco Duque,$^{1}$
and Jonathan Gair$^{1}$
\\
$^{1}$Max Planck Institute for Gravitational Physics (Albert Einstein Institute),
Am M\"uhlenberg 1, D-14476 Potsdam, Germany
}

\date{}
\maketitle

\begin{abstract}
We study the ability of the upcoming Laser Interferometer Space Antenna (LISA) to constrain gas torques acting on extreme-mass-ratio inspirals (EMRIs) when these are embedded in accretion disks, using recently developed relativistic models for the binary–disk interaction. Using a fully Bayesian setup, we find that, contrary to previous forecasts based on Newtonian results, these observations can provide simultaneous estimates of the disk surface density and the accretion rate (or, equivalently, its total luminosity) without the need for an electromagnetic counterpart. Our analysis also indicates that simpler measurement constraints based on the linear-signal (Fisher matrix) approximation are not valid for these systems. For typical EMRI observations, the torque amplitude can be constrained to within $\sim 10\%$, strengthening the prospect of probing accretion physics at (sub)microparsec scales, deep in the strong-field gravity regime and complementing electromagnetic observations. This also strengthens LISA's ability to help answering questions such as how massive black holes grow and coevolve with their host galaxies and, by helping to identify the EMRI's host galaxy through cross-correlation with AGN catalogues, to improve the use of these sources as (dark) sirens for cosmology.
\end{abstract}

\label{firstpage}
%

\section{Introduction}\label{sec:Intro}

One of the most exciting prospects in gravitational-wave (GW) astronomy is the possibility of probing dense astrophysical environments~\citet{Barausse:2014tra}, such as the accretion disks surrounding binary black holes (BHs) in Active Galactic Nuclei (AGN)~\citep{Kocsis:2011dr, Yunes:2011ws, Speri:2022upm, Toubiana:2020drf, Garg:2022nko}. 
Gas in the disk accretes onto the binary components and exerts drag and resonant torques on them~\citep{GoldreichTremaine:1980, 2002ApJ...565.1257T, Tanaka_2024}, perturbing the inspiral trajectory driven by GW emission. 
Although small on an orbital timescale, these effects accumulate over many cycles and leave an observable imprint on the emitted waveform.
If unaccounted for, such environmental effects induce systematic biases in the inferred properties of the binary system, potentially mimicking deviations from General Relativity (GR). On the other hand  
if properly modelled, they give us access to the innermost regions of optically thick accretion disks, complementing electromagnetic observations by probing the nature of accretion flows in the strong-field regime. 
Formation in a dense gaseous environment was suggested as a possible origin for the observed LIGO signals GW190521 \citep{morton2023gw190521binaryblackhole} and GW231123 \citep{Abac_2025}, which are consistent with masses in the supernova pair-instability mass gap ($\sim  50-130  M_\odot$) and high spins \citep{LIGOScientific:2025pvj, Tong:2025wpz, Antonini:2025ilj}, but there was no evidence for the impact of an environment in the short observed signal. The latter is expected to be different for the space-based  mHz GW detector Laser Interferometer Space Antenna (LISA)~\citep{LISA_red_book}, scheduled to launch in 2035.
LISA will observe extreme-mass-ratio inspirals (EMRIs), in which a stellar-mass \textit{secondary} of mass $m_2 \sim 10 - 100\, M_\odot$ spirals into a massive \textit{primary} BH of mass $m_1 \sim 10^5 - 10^7\, M_\odot$, completing up to $\sim 10^5$ orbital cycles within the detector's frequency band~\citep{Babak_2017_EMRIscience, Speri:2026ade}. The resulting signals enable very precise parameter estimation, with relative errors as small as $\sim 10^{-5}$ on spin and eccentricity, making EMRIs ideal probes of small departures from the vacuum-GR driven inspiral, such as those induced by an accretion disk. 

Current electromagnetic observatories keep uncovering new populations of AGNs~\citep{2026A&A...707A.110L},  with $\lesssim 10\%$ of the galaxies in the local Universe ($z \lesssim 0.1$) hosting one, and up to tens of percent near cosmic noon ($z \sim 1$--$2$)~\citep{2014ARA&A..52..589H}, well within LISA's  redshift horizon for EMRIs~\citep{Speri:2026ade}. We therefore expect a non-negligible fraction of LISA EMRIs to be embedded in a gas-rich environment. In fact, the disk itself can actively promote EMRI formation through two processes: the \textit{in-situ} formation of compact objects within the disk plane~\citep{Derdzinski:2022ltb}, and the capture of stellar-mass objects from the surrounding nuclear cluster, whose orbits are progressively damped by repeated disk crossings \citep{Pan_2021_EMRI_in_AGN_formation, Spieksma_2026, Zeng:2026ydj} until they align with the disk and migrate efficiently towards the central BH under gas torques. The latter mechanism has been proposed as the origin of quasi-periodic eruptions (QPEs) observed in X-rays from galactic nuclei~\citep{Linial:2023nqs, Franchini:2023bou, Kejriwal:2024bna, Chakraborty:2025xch, Liu:2026tvy, Pelle:2026ohn, Allievi:2026ycd}.

Many works have investigated the detectability of gas torques on EMRIs with LISA \citep{Kocsis:2011dr, Yunes:2011ws, Speri:2022upm, Cole:2022yzw, Duque:2024mfw, Khalvati_2025}, reaching the consensus that the imprint of a thin, near-Eddington accretion disk on the GW waveform is indeed detectable for realistic disk configurations. 
However, these studies have largely relied on Newtonian models in which, for circular motion, the disk effect reduces to a simple power-law correction to the angular momentum flux
\begin{equation}
    \dot{L} = \dot{L}_{\rm GW} \left( 1 + A\, p^{n_r} \right)\,,
\end{equation}
where $\dot{L}_{\rm GW}$ is the GW contribution, $p$ is the binary's orbital separation, and $n_r$ is a radial scaling index set by the assumed disk model (e.g.,\ $n_r = 8$ for the standard Shakura--Sunyaev $\alpha$-disk~\citep{1973A&A....24..337S}). 
The amplitude $A$ encodes the dependence on disk properties such as the local surface density and accretion rate, which, however, remain degenerate in the waveform. Therefore, although the effect is measurable, the individual disk parameters cannot be disentangled unless an electromagnetic counterpart breaks the degeneracy. Recent work has shown that orbital eccentricity can help lift these degeneracies~\citep{Duque:2024mfw, 2025MNRAS.543..565F}. Another source of uncertainty is the stochastic variability of gas torques observed in hydrodynamic simulations~\citep{Derdzinski:2020wlw, Zwick:2021dlg, Derdzinski:2025cql}, which can bias the inferred disk parameters~\citep{Copparoni:2025jhq} and, if sufficiently strong, invalidate torque estimates based on linear theory~\citep{Wu:2023qeh, Garg:2026jrv}. 

More recently, different authors have developed fully relativistic models of the EMRI--disk interaction for circular, equatorial orbits in simplified disk configurations~\citep{Duque:2025yfm, Hegade_model, HegadeKR:2025rpr, Dyson:2026ddd}. 
These works show that, similarly to what was concluded for other environments~\citep{Cardoso:2021wlq, Cardoso:2022whc, Duque:2023seg, Dyson:2025dlj, Li:2025ffh, Vicente:2025gsg}, general-relativistic corrections can amplify the magnitude of the gas torques by up to an order of magnitude compared to Newtonian predictions.
They also show that the radial dependence of the torque is no longer a simple power-law and depends non-trivially on the primary spin and disk profile, suggesting that even circular EMRI observations could, in principle, yield more information about the disk structure and the nature of the gas flow than previously thought.

This is precisely what we investigate in this work. 
Building on~\citep{Duque:2025yfm, Hegade_model, HegadeKR:2025rpr}, we implement a fully relativistic model of the gas forces exerted on a circular, equatorial EMRI embedded in a relativistic accretion disk into a state-of-the-art EMRI waveform model, and perform a fully Bayesian analysis to determine how precisely LISA observations could measure internal disk properties such as the local density and accretion rate. Remarkably, we find that for astrophysically plausible configurations, this is possible without the need for an electromagnetic counterpart, substantially improving on previous forecasts based on Newtonian, power-law models.

Unless explicitly stated, we adopt geometric units in which $G= c= 1$ and distances are measured in terms of the primary mass $m_1$.

\section{Setup and methods}\label{sec:methods} 


\subsection{Disk and gas torque model}


For the accretion disk, we consider the Novikov--Thorne profile~\citep{1973blho.conf..343N}.
This is the relativistic extension of the fiducial Shakura--Sunyaev $\alpha$-disk model~\citep{1973A&A....24..337S}, and describes a thin, optically thick, radiatively efficient flow around a Kerr BH in an axisymmetric, stationary configuration. 
The disk lies in the equatorial plane of the primary and extends from the innermost stable circular orbit (ISCO) outwards, with matter on nearly geodesic circular orbits; following the standard prescription, disk torques are assumed to vanish at the ISCO. 
All disk quantities are vertically averaged over the scale height $H$, a standard approximation for geometrically thin disks, where the aspect ratio $h \equiv H/r \ll 1$ everywhere~\citep{Abramowicz:2011xu}.

The disk model is fully characterised by four parameters: the primary mass $M_1$ and dimensionless spin $a$; the mass accretion rate, conventionally expressed in units of the Eddington rate as $f_{\rm Edd} \equiv \dot{M_1}/\dot{M}_{\rm Edd}$, with $\dot{M}_{\rm Edd} = 4\pi G M_1 / \eta \, c\, \kappa_{\rm es} \sim 2\times10^{-2}\, M_\odot {\rm yr}^{-1}\  (M_1/10^6 M_\odot)\,\left(0.1/\eta\right)  $, with $\kappa_{\rm es}$ is the electron-scattering opacity and $\eta$ the radiative efficiency in converting rest-mass energy into radiation (typically $\eta \sim 0.1$); and the dimensionless viscosity parameter $\alpha$, which links the shear stress to the total pressure via $T_{r\phi} = \alpha\, P$, where $T_{\mu \nu}$ is the energy-momentum tensor of the disk fluid. For astrophysically motivated AGN thin disks, we consider Eddington ratios in the range $f_{\rm Edd} \sim 10^{-2} - 10^{-1}$ and viscosities $\alpha \sim 0.01$--$0.1$, consistent with values inferred from electromagnetic observations of AGN~\citep{Kollmeier_2006, Kelly_2010} and from magnetohydrodynamic simulations of accretion flows~\citep{Jiang__2019_AGN_sim_accrate}. The equations for the disk density $\Sigma_{\rm NK}$ and the aspect ratio $h_{\rm NK}$ are the following:
\begin{align}
&\Sigma_{\rm NK} (r, a) = \Sigma_0 \  r^{3/2} F(r, a)\, , \\
& h_{\rm NK} (r, a) = 3 h_0 \left ( \frac{10 M_1}{r}\right ) G(r,a) \, ,
\label{eq:sigma_eq}
\end{align}
where $F$ and $G$ are characteristic functions of the Novikov-Thorne disk profile ~\citep{1973blho.conf..343N, Abramowicz:2011xu}, with values of $\mathcal{O}(1-10)$ at $r\sim 10M_1$. We relate the characteristic amplitude of the surface density profile $\Sigma_0$ and aspect ratio with the viscosity parameter $\alpha$ and the Eddington rate $f_{\rm Edd}$ through
\begin{align}
\Sigma_0 &= 1.6\times 10^5 \left(\frac{0.1}{\alpha} \right)  \left(\frac{0.1}{f_{\rm Edd}}\right) \ \, [\rm Kg/m^2] \label{eq:Sigma0_h0_eq_1}  \, , \\
h_0 &= 1.5 \ f_{\rm Edd} . \label{eq:Sigma0_h0_eq_2}
\end{align}
%
Choosing $\alpha \sim 0.1$, $f_{\mathrm{Edd}} \sim 0.1$ the surface density ranges between $\Sigma_{\rm NK} \in [1.1, 4.4] \times 10^{7} \  \mathrm{Kg\,/ m^{2}}$, monotonically decreasing as a function of the primary spin parameter $a$. While the aspect ratio spans $h_{\rm NK} \in [0.045, 0.19]$, monotonically increasing with the spin parameter. These intervals correspond to $a \in [0, 0.99]$. Adopting Eq.~\eqref{eq:Sigma0_h0_eq_1} and \eqref{eq:Sigma0_h0_eq_2}, one can, for a given primary mass, convert from $(\Sigma_0,\, h_0)$ to $(\alpha,  \,f_{\rm Edd})$ and vice-versa. 

For the relativistic torque exerted on the secondary, we follow the framework of~\citet{Duque:2025yfm, Hegade_model, HegadeKR:2025rpr}. 
These works treat the disk as a collection of test particles following circular geodesics in the equatorial plane, and compute the transfer of angular momentum between a circular, equatorial EMRI and such particles. This corresponds to modelling the disk as 2D and pressureless, and to neglecting the co-orbital region around the secondary. Newtonian hydrodynamic simulations~\citep{Masset_2002, Thin_disk_sim, Derdzinski:2025cql},  have shown that the 2D treatment is accurate as long as the secondary's Hill radius, $r_{\rm H} \equiv r\,(m_2/3 m_1)^{1/3}$, remains smaller than the disk scale height $H$, a condition satisfied throughout this work. 
Pressure effects, in turn, become relevant only at distances from the secondary comparable to the mean free path of disk particles, which in a thin viscous disk is of order of the scale height $\sim H$. Pressure is therefore negligible for the scales governing the net torque, except near the secondary itself, where they regularise the total torque, as explained below~\citep{1993ApJ...419..155A, Ward:1997, 2012ARA&A..50..211K, Duque:2025yfm}. Co-orbital (horseshoe) torques are highly sensitive to entropy and vorticity gradients across the horseshoe region and are expected to be subdominant in the $r_{\rm H} \ll H$ limit~\citep{Derdzinski:2025cql, Tanaka_2024}. Finally, the assumption of circular, equatorial motion is well-justified for EMRIs formed in this channel as disk interactions efficiently damp eccentricity and inclination, so that by the time the EMRI enters the LISA band the orbit is expected to be nearly circular and aligned with the disk~\citep{Spieksma_2026, Zeng:2026ydj}.

Under these approximations, the problem reduces to computing the secular angular-momentum exchange between the EMRI and the test particles of the disk. 
Since this exchange is much smaller than the EMRI's orbital angular momentum, the inspiral evolves adiabatically and the torque can be obtained by averaging over many orbital cycles~\citep{Barack:2018yvs, Pound:2021qin, Hughes:2021exa}. 
The only particles that contribute to the net torque in this average are those at \textit{Lindblad resonances}~\citep{GoldreichTremaine:1980, 2002ApJ...565.1257T}, where the azimuthal mode-$m$ perturbation of the secondary resonates with the radial epicyclic motion of a slightly perturbed disk particle. 
In the relativistic, Kerr regime, this condition reads
\begin{equation}
    m\, \Omega_\phi^{\rm EMRI} = m\, \Omega_\phi^{\rm disk} \pm \Omega_r^{\rm disk}\,,
\end{equation}
where $\Omega_\phi$ and $\Omega_r$ are the azimuthal and radial epicyclic frequencies of a circular equatorial orbit, and the upper and lower signs label respectively the outer and inner Lindblad resonances, located outside and inside the secondary's orbit.
These produce torques of opposite sign, whose partial cancellation sets the overall magnitude and sign of the disk's effect on the inspiral.

The main difference between the treatments of~\citet{Duque:2025yfm} and~\citet{Hegade_model, HegadeKR:2025rpr} lies in how the resonance amplitudes are computed. 
Both works start from a Hamiltonian formulation of the disk--secondary interaction. ~\citet{Duque:2025yfm} follows~\citet{Hirata_1, Hirata_2}, who related the torque amplitudes to the asymptotic behaviour at large distances and at the primary BH horizon of solutions to the Teukolsky equation~\citep{1973ApJ...185..635T}, which is the master equation governing gravitational perturbations of Kerr.~\citet{Hegade_model, HegadeKR:2025rpr} instead compute the singular field of the secondary directly, and use it to evaluate the perturbed gravitational potential felt by a nearby disk particle. This is possible because the Lindblad resonances lie in the near-zone of the secondary, where the singular field provides an accurate local description of the perturbation. We have verified that the resulting torque amplitudes agree to within $\lesssim 1\%$, consistent with numerical error. A further subtlety is that, in a strictly pressureless disk, the sum of Lindblad-resonance contributions formally diverges as resonances accumulate towards the secondary. Following~\citet{Duque:2025yfm}, we regularise this divergence by introducing a phenomenological cutoff that smoothly suppresses the contribution of resonances located within $\sim H$ of the secondary. This corresponds to the standard \textit{torque cutoff} of planetary-migration theory~\citep{GoldreichTremaine:1980,Artymowicz:1993,Ward:1997}. In the Newtonian limit and for a power-law disk profile, this framework reduces to the classical Goldreich--Tremaine result~\citep{ GoldreichTremaine:1980}, originally derived in the context of protoplanetary migration. The relativistic framework retains this resonant structure, but modifies the resonance locations, the radial density profile, and the response function at each resonance. Together, these contribute to the order-of-magnitude enhancement of the net torque in the strong-field regime, as illustrated in Fig.~1 of~\citet{Duque:2025yfm}. For the disk models considered here, outer resonances systematically dominate over inner ones, so that the net gas torque accelerates the inspiral. Interestingly, relativistic effects can reverse the sign of this torque close to the ISCO~\citep{Duque:2025yfm}, but this reversal occurs deep in the strong-field regime where GW emission dominates the evolution, so it is not expected to leave a significant observational imprint. The reversal radius also moves progressively closer to the ISCO for larger primary spin, and accreting BHs are generally expected to spin rapidly, close to the Thorne limit $a \simeq 0.998$~\citep{Thorne:1974}, though massive BHs can grow in mass significantly while keeping a low spin if accretion is chaotic~\citep{Chaotic_accretion}. 
In this work, we will focus on spins of $a \geq 0.7$. 


\subsection{Integration into EMRI waveforms}



For the EMRI waveforms we use the \texttt{FastEMRIWaveform} (\texttt{few}) package \citep{FEW_paper}, a modular, GPU-accelerated Python framework that generates EMRI waveforms in hundreds of milliseconds. Its latest release implements a fully relativistic adiabatic model (accurate to linear order in the mass ratio) for eccentric, equatorial EMRIs in a Kerr background, valid up to very large orbital radii \citep{Chapman_Bird_2025} 
We restrict to circular, equatorial orbits and take both the secondary and the disk to be prograde with respect to the primary. This is expected from angular momentum alignment mechanisms like the Bardeen--Petterson effect~\citep{1975ApJ...195L..65B}, and disk torques also align the EMRI
orbital angular momentum effectively before the system enters the LISA band~\citep{Spieksma_2026, Zeng:2026ydj}.

\texttt{few}'s modularity lets us add the disk-torque correction directly to
the angular momentum flux,
\begin{equation}
    \dot{L} \;=\; \dot{L}_{\rm GW} + \dot{L}_{\rm disk}\,,
\end{equation}
with the disk contribution factorizing linearly in the reference surface
density,
\begin{equation}
    \dot{L}_{\rm disk}(r, a, \Sigma_0, h_0)
    \;=\; \Sigma_0\, \tilde{\dot{L}}_{\rm disk}(r, a, h_0)\,.
\end{equation}
We evolve the inspiral trajectory under this modified flux balance and generate the waveform along the resulting trajectory assuming a Kerr background. 

The torque model is too slow for on-the-fly evaluation during Bayesian
parameter estimation, so we precompute a dense grid over $(r, a, h_0)$
and interpolate during inference. The grid is uniform in spin $a \in [0, 0.99]$, aspect ratio $h_0 \in [0.01, 0.5]$, and radius $r \in [r_{\rm ISCO}(a) + 2, 30]\,M$, with $600 \times 200 \times 200$
points. To validate the grid is sufficiently dense for the interpolation to be accurate enough for the precision of EMRI observations, we compute the integrated relative difference
\begin{equation}
    {\rm IRD} = \int _{r_{\rm ISCO}(a) + 2}^{30} \frac{(T_{\rm int}(r) - T_{\rm true}(r))}{T_{\rm true}(r)} \ dr
\end{equation}
where $T_{\rm int}$ is the interpolated torque and $T_{\rm true}$ is computed directly. Across $10^{3}$ pairs
$(\Sigma_0, h_0)$ drawn away from the grid knots we find median ${\rm IRD} = 2.8 \times 10^{-6}$ and a max value of 
${\rm IRD} = 1.4 \times 10^{-3}$, corresponding to a dephasing of $< 1 $~rad, below the rule of thumb threshold for any observational impact at the signal-to-noise ratios we will consider. 

We set the disk torque to zero for $r < r_{\rm ISCO}(a) + 2$, since the
Novikov--Thorne disk model is not accurate in this regime~\citep{Potter_2021, Duque:2025yfm}. The cutoff does not affect our results since at
these radii the GW torque exceeds the disk torque by $7$--$8$ orders of
magnitude, compared to $\sim$ $2$--$4$ orders of magnitude in the early inspiral (see Fig.~1 in ~\citet{Duque:2025yfm}). 

The vacuum flux $\dot{L}_{\rm GW}$ depends on the primary mass $M_1$, the secondary mass $M_2$, the primary spin $a$, the orbital separation $p$, and the initial azimuthal phase $\Phi_0$. The full waveform adds two angle
pairs: $(\theta_k, \phi_k)$, the orientation of the primary's spin, and $(\theta_s, \phi_s)$, the sky location in the Solar System barycentre frame, together with the luminosity distance $d_L$. We call
$(M_1, M_2, a, p, \Phi_0)$ the \emph{intrinsic} parameters and $(\theta_k, \phi_k, \theta_s, \phi_s, d_L)$ the \emph{extrinsic} parameters. In this work we assume to observe the last 4 years of the signal, ending at plunge. These systems are the most likely to be detected, due to the large contribution to the total signal-to-noise ratio (SNR) from the close-to-plunge segment of the inspiral~\citep{Speri:2026ade}.

\subsection{Parameter Estimation Framework}


\begin{table*}
\centering
\setlength{\tabcolsep}{6pt}
\renewcommand{\arraystretch}{1.15}
\begin{tabular}{l c c c c c}
\toprule
\textbf{Parameter} & \textbf{Strong} & \textbf{Intermediate ($a=0.7$)} & \textbf{Intermediate} & \textbf{Massive} & \textbf{Weak} \\
 & ($\Delta\Phi = 105\,\mathrm{rad}$) & ($\Delta\Phi = 88\,\mathrm{rad}$) & ($\Delta\Phi = 52\,\mathrm{rad}$) & ($\Delta\Phi = 11\,\mathrm{rad}$) & ($\Delta\Phi = 2.6\,\mathrm{rad}$) \\
\midrule
$M_1$ $[M_\odot]$              & $2.5\times 10^5$ & $2.5\times 10^5$ & $2.5\times 10^5$ & $10^6$           & $2.5\times 10^5$ \\
$M_2$ $[M_\odot]$              & $10$             & $10$             & $10$             & $50$             & $10$             \\
$a$                            & $0.9$            & $0.7$            & $0.9$            & $0.9$            & $0.9$            \\
$\Phi_0$ $[\mathrm{rad}]$      & $1.0$            & $1.0$            & $1.0$            & $1.0$            & $1.0$            \\
$\theta_k$ $[\mathrm{rad}]$    & $1.05$           & $1.05$           & $1.05$           & $1.05$           & $1.05$           \\
$\phi_k$ $[\mathrm{rad}]$      & $1.05$           & $1.05$           & $1.05$           & $1.05$           & $1.05$           \\
$\theta_s$ $[\mathrm{rad}]$    & $0.52$           & $0.52$           & $0.52$           & $0.52$           & $0.52$           \\
$\phi_s$ $[\mathrm{rad}]$      & $0.52$           & $0.52$           & $0.52$           & $0.52$           & $0.52$           \\
$p_0$ $[M]$                    & $20.87$          & $21.09$          & $20.87$          & $15.49$          & $22.09$          \\
$d_L$ $[\mathrm{Gpc}]$         & $2.12$           & $2.07$           & $2.12$           & $6.09$           & $2.76$           \\
$z$                            & $0.38$           & $0.37$           & $0.38$           & $0.91$           & $0.49$           \\
\midrule
$\alpha$                       & 0.05              & 0.09              & 0.09              & 0.09              & 0.24              \\
$f_{\rm Edd}$                  & 0.03              & 0.03              & 0.03              & 0.03              & 0.07             \\
$\Sigma_0$ $[\mathrm{kg/m^2}]$ & $10^6$           & $5\times 10^5$   & $5\times 10^5$   & $5\times 10^5$   & $10^5$           \\
$h_0$                          & $0.05$           & $0.05$           & $0.05$           & $0.05$           & $0.10$           \\
\bottomrule
\end{tabular}
\caption{The five EMRI configurations adopted in this work, listed in
decreasing order of the inspiral dephasing $\Delta\Phi$ induced by the
relativistic disk torques. Masses are quoted in the detector frame. The luminosity distance of each system is chosen to give a signal-to-noise
ratio of $50$. Redshifts are computed assuming Planck~2018 cosmology~\citep{Planck:2018vyg} }
\label{tab:Systems_summary}
\end{table*}

We perform a fully Bayesian analysis to assess how well disk torques can be
constrained for typical EMRI systems. Our goal is to obtain the posterior
distribution for the EMRI parameters $\lambda$ given a signal $s$,
\begin{equation}
p(\lambda \mid s) \;=\; \frac{\mathcal{L}(s \mid \lambda)\,\pi(\lambda)}{\mathcal{Z}}\,,
\end{equation}
where Bayes' theorem expresses the posterior in terms of the likelihood
$\mathcal{L}(s \mid \lambda)$, the prior on the EMRI parameters $\pi(\lambda)$,
and the evidence $\mathcal{Z} = \int \mathcal{L}(s \mid \lambda)\,\pi(\lambda)\,d\lambda$.
Since we are interested only in parameter estimation and not in model
comparison, $\mathcal{Z}$ acts as a $\lambda$-independent normalization
constant and can be neglected.

We sample the posterior with the
Markov-Chain Monte Carlo sampler \texttt{eryn}~\citep{Eryn_paper}, using 36
walkers and a single temperature. We initialize the walkers close to injection, typically within the main mode of the posterior distribution. For this reason, we do not need a parallel tempering scheme to explore the rest of parameter space and discard secondary modes. For every run, we obtain enough samples such that the chain length is larger than $50\,\bar{\tau}$, where $\bar{\tau}$ is the mean autocorrelation time across walkers~\citep{Hogg_2018}. We adopt a standard Gaussian likelihood,
\begin{equation}
    \mathcal{L}(\lambda) \;\propto\; \exp\!\left\{
    -\tfrac{1}{2}\, \langle s - h(\lambda) \mid s - h(\lambda) \rangle
    \right\},
\end{equation}
where $s$ is the injected signal, $h(\lambda)$ the waveform model, and
$\lambda$ the full parameter set (intrinsic, extrinsic, and environmental).
The inner product is
\begin{equation}
    \langle a \mid b \rangle
    \;=\; 4\,\mathrm{Re} \int_0^\infty
    \frac{\tilde{a}^*(f)\, \tilde{b}(f)}{S_n(f)}\, \mathrm{d}f,
\end{equation}
with $\tilde{a}(f)$ and $\tilde{b}(f)$ the Fourier transforms of $a(t)$
and $b(t)$, and $S_n(f)$ the one-sided power spectral density (PSD) of LISA. A particular noise realization shifts the posterior maximum away from
the injected values, without appreciably broadening their distribution, corresponding to a loss of accuracy rather than precision.  For this reason, we are limit this study to noiseless realizations. Real LISA observations will exhibit non-Gaussian noise, instrumental glitches, and data gaps from scheduled maintenance or 
unplanned interruptions, all of which complicate the likelihood and could induce biases in results computed from simplified likelihood models ~\citep{Burke:2025bun, Burke:2025fvl, Boumerdassi:2025gvf}. 

LISA data analysis for EMRIs will most likely be carried out in two stages: a search phase
~\citep{Cole:2025sqo,Strub:2025dfs,Speri:2025ucn} that locates the
approximate maximum of the posterior, followed by a parameter-estimation phase that determines a more accurate measurement of the parameters and their uncertainties. We assume that our signals can be successfully recovered in the search phase and focus only on parameter estimation. This is plausible even in the presence of disk torques,
since for the mass ratios considered most of the SNR accumulates during
the final $\sim 6$~months of the inspiral~\citep{Copparoni:2025vty},
where disk effects are largely negligible. We would therefore expect that a search tuned to find vacuum EMRI signals would also find EMRIs being influenced by this kind of environmental effect. For the parameter estimation phase, we take priors on the intrinsic EMRI parameters that are uniform over a $1.5\%$ window centered on the injected value, while priors on the extrinsic parameters are wider, spanning up to 
$\sim 60\%$ of the physically allowed range, to ensure that the posterior
is well contained within the prior.

As we will see below, the EMRI parameter space exhibits strong
correlations between the two environmental
parameters ($\Sigma_0, h_0$). This
makes sampling computationally heavy (runs can take up to $\sim 1$ week to
satisfy the convergence criterion defined above). To speed up inference, we spread the initial walkers along directions indicated by the Fisher Information matrix
\begin{equation}
    \Gamma_{ij} = \left\langle
    \frac{\partial \ell}{\partial \lambda_i}\ \middle\vert
    \frac{\partial \ell}{\partial \lambda_j}\right\rangle
\end{equation}
which approximates the curvature of the log-likelihood $\ell = \log \mathcal{L}$ at the best-fit point under the linear-signal approximation, such that the inverse of the Fisher Information matrix is the covariance matrix of the parameter posterior. As we show below, this approximation is \emph{not} accurate even at high SNR, which prevents us from carrying out an exhaustive study of the measurability of the relativistic disk torques across the full parameter space, but it remains useful for setting the initial state of the sampler. We initialize the sampler along the directions indicated by the Fisher matrix, reducing all its contours to smaller scale with the Fisher matrix prediction.  

The Fisher matrix also identifies the most strongly correlated parameter combinations, which suggests reparametrisations that improve MCMC convergence.
We find a strong correlation between $\Sigma_0$ and $h_0^2$ that becomes
a full degeneracy in the Newtonian limit since the disk torque is linear in $A$, as highlighted in \citet{Speri:2022upm}.
The data therefore constrain the combination $A \equiv \Sigma_0/h_0^2$ much
better than either parameter individually. To (quasi-)diagonalize the covariance
along this degenerate direction, we sample in
$u = \ln(\Sigma_0/h_0^2)$ and $v = \ln(h_0\,\Sigma_0^2)$, and adopt
log-uniform priors on $\Sigma_0$ and $h_0$ in physical space, transformed
consistently to $(u,v)$.





\subsection{Fiducial sources}


Since our fully Bayesian setup is computationally expensive, we restrict the
analysis to a few representative EMRI--disk systems. We pick only
two mass combinations, $(M_1, M_2) = (10^6,\, 50)\, M_\odot$ and $(2.5\times 10^5,\, 10)\, M_\odot$ in the detector frame, and two primary
spins, $a = 0.9$ and $a = 0.7$. The more massive EMRI sits in the bulk of the LISA sensitivity band, whereas the less massive one chirps more rapidly, meaning
for a fixed observation window prior to plunge, it evolves from a
larger orbital separation, where the disk torques are stronger.
Then, we consider the five EMRI-disk configurations listed in Tab. \ref{tab:Systems_summary}, varying the strength of the disk torques. 

We label four of the setups as: \textit{strong}, \textit{intermediate $a=0.7$}, \textit{intermediate},  and \textit{weak}, based on the induced dephasing $\Delta \Phi$ of the waveform at plunge with respect to the waveform from the vacuum case (and the $a=0.7$ to distinguish from the other cases which have $a=0.9$). The fifth setup is denoted \textit{massive} as it uses a heavier primary black hole mass, while all other cases use the lighter mass. The stronger the environmental effect, the larger the dephasing accumulated during the evolution and the better constraints we expect. Note that the
reduction of the primary spin from $a = 0.9$ to $a = 0.7$ at
otherwise identical disk parameters shifts the accumulated dephasing
from $52$ to $88\,\mathrm{rad}$. This is because a smaller spin pushes the ISCO outward so, for a fixed
observation time, the EMRI then starts at a larger radius, where the disk
torques are stronger, and accumulates a larger dephasing before
plunge.

For each system we assume the nominal 4-year LISA observation prior to plunge, and we
place the source at a luminosity distance such that $\mathrm{SNR} = 50$.  All sources share the same sky position and orientation: $(\theta_k, \phi_k, \theta_s, \phi_s) = (\pi /3, \pi /3, \pi/6, \pi/6)$.
Since the extrinsic parameters are weakly correlated with the intrinsic and
environmental parameters, our results are 
insensitive to this choice. 


\section{Results}\label{sec:results}

\begin{figure*}
\includegraphics[width=0.99\linewidth]{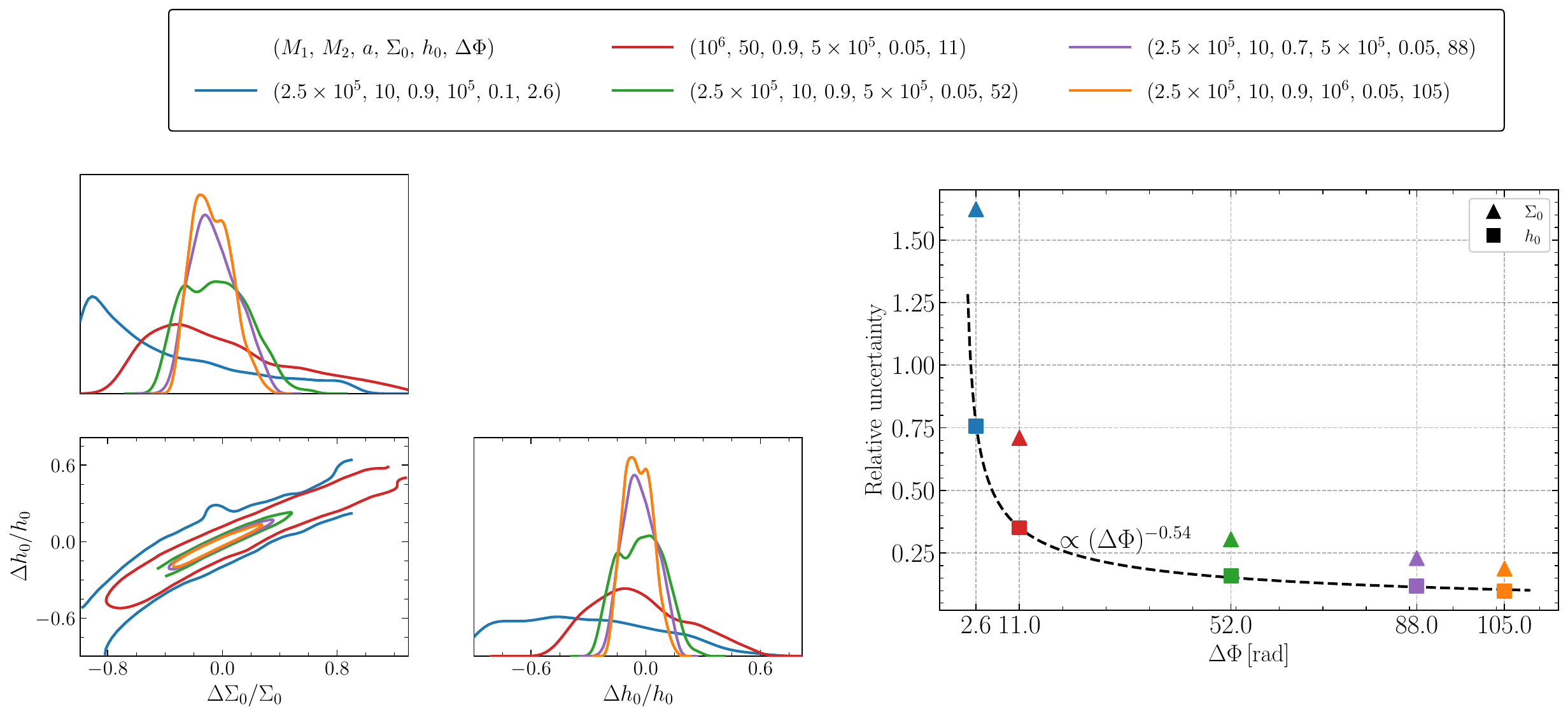}
\caption{\textit{Left panel}: Marginal 2D posteriors of the environment
parameters in units of the relative deviation from the injected value,
$(X - X_{\rm true})/X_{\rm true}$. Colors correspond to the different
torque models, ordered by increasing dephasing.
\textit{Right panel}: Relative uncertainty (defined by Eq.~(\ref{eq:rel_unc})) 
as a function of the dephasing. Colors are
consistent with the left panel. Triangles and squares indicate the
relative uncertainties on $\Sigma_0$ and $h_0$, respectively. The
dotted line shows a fit of the relative uncertainty versus dephasing.}
\label{fig:env_summary}
\end{figure*}

\subsection{Measurement uncertanties}

Our main results are illustrated in Fig.~\ref{fig:env_summary}. The
left panel shows the marginal 2D posteriors of the environment
parameters in units of the relative deviation from the injected value. 
A posterior of width $|\Delta\Sigma/\Sigma_0| \sim 1$ therefore corresponds
to an environment that is indistinguishable from vacuum within the statistical uncertainty.
The right panel shows the relative uncertainty $U$ (defined in
Eq.~\ref{eq:rel_unc} below) as a function of the accumulated dephasing
$\Delta\Phi$ for each of the five configurations of
Table~\ref{tab:Systems_summary}. The dashed
line is a  power-law fit to the $h_0$ points, also discussed
quantitatively below. For illustration, we also show in
Fig. \ref{fig:Full_Strong_case} of Appendix~\ref{sec:App_FullCorner} the full corner
plot for the fiducial \textit{strong} system of Table~\ref{tab:Systems_summary}.

The posteriors in the left panel exhibit a strong correlation between
$\Sigma_0$ and $h_0$, consistent with the direction predicted by the
Fisher information matrix. As mentioned above, these two parameters
are completely degenerate in the Newtonian limit, so previous
analyses could only constrain the combination
$A \equiv \Sigma_0/h_0^2$~\citep{Speri:2022upm, Khalvati_2025}. The relativistic torque model we employed lifts the degeneracy. When the torque is strong enough
to induce a dephasing $\Delta\Phi \gtrsim 10\,\mathrm{rad}$, $\Sigma_0$
and $h_0$ become separately constrained, marking a qualitative
departure from Newtonian-based analyses. For the \textit{weak} configuration, the
posterior collapses onto the $A = \mathrm{const.}$ ridge and only $A$
remains informative. In this case, the posteriors broaden
to the point of being degenerate with the prior in $h_0$. We compare our results with~\cite{Khalvati_2025}, finding for a similar system configuration a relative standard deviation $\sigma_{A}/A \sim 8 \%$, consistent with $\sim 16\%$ from~\cite{Khalvati_2025}. The increased precision of our results is a consequence of the amplification of relativistic torque with respect to the Newtonian one on the same system \cite{Duque:2025yfm}. When the dephasing is significant, the marginal posterior on the parameter $u$ has a Gaussian-like shape, but exhibits non-Gaussian features for weaker torques. We find a similar behavior when we fix the vacuum EMRI parameters and sample only on the environmental parameter combinations $u$ and $v$. As expected, the posterior is more constrained with respect to the complete MCMC runs. The correlation with the other parameters of the EMRI system leads to a factor of $\sim 5$ and $\sim 3$ decrease in the precision of the environmental combination $u$ and $v$ respectively. We notice that for the \textit{strong} configuration the Fisher matrix approximation represents perfectly the posterior distribution of the extrinsic parameters (not correlated with the environment), it performs well for the intrinsic parameters, catching the spread of the marginal distributions and providing the correct correlation directions. However, it performs badly on the environmental parameters, predicting a broader distribution. The observed behaviour is most likely due to the strong degeneracy of the environmental parameters, that breaks the linear signal approximation even for large dephasings.  For each configuration, we quantify the constraining power through the relative uncertainty
\begin{equation}
    U \;\equiv\; \frac{Q_{95} - Q_{5}}{2\,Q_{50}} \quad ,
    \label{eq:rel_unc}
\end{equation}
where $Q_{50}$ is the posterior median and $Q_{5,95}$ are the
extremes of the $5^{th}$ and $95^{th}$ quantile of the distribution credible interval. For the fiducial
\textit{strong} configuration, we recover the injected parameters at
$U \sim 18\%$ for $\Sigma_0$ and $U \sim 9\%$ for $h_0$.

The right panel of Fig.~\ref{fig:env_summary} shows that the relative uncertainty in both $\Sigma_0$  and $h_0$ monotonically decreases with  
$\Delta\Phi$, as expected. This happens even for the \textit{massive} configuration ($M_1 = 10^6\,M_\odot$,
$M_2 = 50\,M_\odot$) and the \textit{intermediate} ($a=0.7$) configuration which have distinct masses and orbital evolution. We fitted this variation to a power-law
$U = A\,(\Delta\Phi)^{\gamma}$,  obtaining
\begin{equation}
\begin{aligned}
A_{h_0}     &= 1.28 \pm 0.08, & \gamma_{h_0}     &= -0.54 \pm 0.02, \\
A_{\Sigma_0}&= 2.83 \pm 0.14, & \gamma_{\Sigma_0}&= -0.57 \pm 0.01,
\end{aligned}
\label{eq:scaling}
\end{equation}
so the slopes are consistent within their uncertainties. The fit passes through each point across two orders of magnitude for $\Delta \Phi$, therefore giving a computationally cheap forecasting of the measurement uncertainties for the disk parameters. In Appendix~\ref{sec:App_Dephasing}, we provide the dephasing for more EMRI-disk systems. 

The recovered scaling $\gamma \approx -1/2$ is shallower than the naive Fisher expectation.  In the Gaussian linear-signal
approximation, the relative uncertainty at fixed SNR scales as
$U \propto (\mathrm{SNR}|\partial h/\partial\theta|)^{-1}$. For small disk effects,  the phase response is approximately linear in the disk parameters,
$\Delta\Phi \propto \theta$, so we expect
$U \propto (\Delta\Phi)^{-1}$. One of the possible explanations for the
$\gamma \simeq -0.5$ scaling is the departure of the signals from the
Gaussian regime, consistently with the failure of the Fisher
approximation discussed above. However, there are other possibilities worth mentioning. Firstly, the sampling is done in the $(u,v)$ space, and the relative uncertainty for $u$ and $v$ as a function of dephasing $\Delta \Phi$ scales as a power laws, with slope $\gamma_u \sim -1$ and $\gamma_v \sim -1/2$ respectively; the power law slope on $v$ may reflect the departure from the Gaussian regime. The relative uncertainty on $v$ is larger than the one on $u$ and therefore dominates the trend on the physical space quantities, $(\Sigma_0, h_0)$. At the same time, in the set of systems chosen we are also varying parameters other than the environmental ones and we expect these parameters to vary the prefactor in the uncertainty scaling. For this reason, the $\gamma \sim -1/2$ slope observed could also be the result of fitting points from $\gamma \sim -1 $ power laws with different amplitudes.

An example of the complete posterior, including all free parameters
of the model, is shown in Fig. \ref{fig:Full_Strong_case} for the
\textit{strong} configuration. The combination
$u = \ln(\Sigma_0/h_0^2)$ exhibits strong correlations with all the
intrinsic parameters and a weaker correlation with the polar sky
angle $\phi_s$. Since $u$ is directly related to the torque amplitude
$u \sim \ln A$, these correlations are expected. As one increases $A$,
the inspiral is faster, which can be mimicked by a larger
initial separation $p_0$ over the same evolution time, by a larger
secondary mass $M_2$, or by a smaller primary mass $M_1$, explaining
the signs of the corresponding correlations. The anti-correlation
with the primary spin $a$ has the same origin. For a prograde orbit,
a smaller spin places the ISCO at larger radii and shortens the
strong-field portion of the inspiral, partially compensating for a
stronger torque.

\subsection{Probing AGN physics and multimessenger astronomy}\label{sec:EM_measurement}

\begin{figure}
\includegraphics[width=0.99\linewidth]{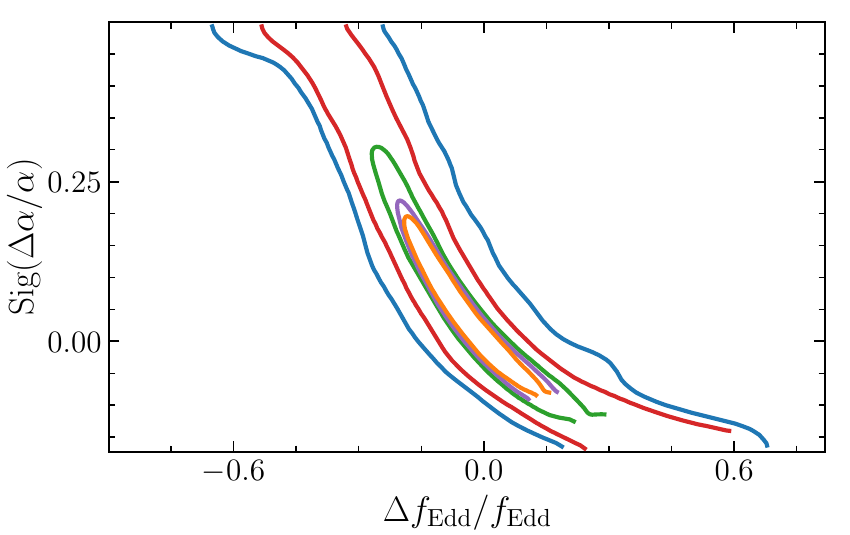}
\caption{The 2D contours of the relative deviation with respect to the true values for the viscosity parameter $\alpha$ and the accretion rate $f_{\mathrm{Edd}}.$ the relative deviation for $\alpha$ has been transformed through a sigmoid function for visualization purposes.}
\label{fig:f_edd_alpha_plot}
\end{figure}

Since with the relativistic torque model we are able to measure both $\Sigma_0$ and $h_0$, this implies that we can measure the viscosity
parameter $\alpha$ and the Eddington ratio $f_{\rm Edd}$ from
GWs alone, with no need for an electromagnetic
counterpart. Using the mapping in Eq.~\eqref{eq:sigma_eq}, the
\textit{intermediate} EMRI-disk configuration of
Table~\ref{tab:Systems_summary} yields
$f_{\rm Edd} = 0.033^{+0.006}_{-0.006}$ and
$\alpha = 0.10^{+0.09}_{-0.04}$, or $\sim 19 \%$ and $\sim  60\%$ relative uncertainty, respectively. The corresponding 2D marginal
posteriors in the $(\alpha,f_{\rm Edd})$ space for all EMRI-disk systems studied is shown in
Fig.~\ref{fig:f_edd_alpha_plot}. These results imply that EMRIs observed with LISA can directly
probe the nature of gas flow in the strong-field regime. A GW measurement of both $\alpha$ and $f_{\rm Edd}$ could
discriminate between different accretion-disk models and yield evidence of super-Eddington
accretion.\footnote{We note that for $f_{\rm Edd} \gtrsim 1$ the
thin-disk approximation $h \lesssim 0.1$ breaks down. Pressure
effects and advection become relevant~\citep{Slim_acc_disks, 2011arXiv1108.0396S, Kao:2025gxz}, and both the disk model and
the corresponding torque prescription used in this work would need
to be modified.}

This result is a substantial improvement over the Newtonian framework
of~\citet{Speri:2022upm}, under which assumptions, as mentioned above, it is only possible to constrain the
torque amplitude $A = \Sigma_0/h_0^2 \propto 1/(\alpha f_{\rm Edd}^3)$.
Breaking the $(\alpha, f_{\rm Edd})$ degeneracy in that
setting requires an electromagnetic counterpart, which allows the disk bolometric
luminosity to be measured from the electromagnetic signal directly to obtain
$f_{\rm Edd}$, from which $\alpha$ then follows using the GW measurement. This is, however, a
highly optimistic scenario. For the EMRI systems we studied, the LISA sky-localization error
$\Delta\Omega \in [0.15, 1.7] \ \rm deg^2$. At the assumed redshifts, that sky volume contains $\mathcal{O}(10^2-10^3)$ candidate galaxies~\cite{Laghi:2021pqk, laghi2021gravitationalwavecosmologyemris, 2023MNRAS.519.5962L}, within which one would need to identify a
distinctive modulation in the electromagnetic emission,
originating on $10M_1 \sim 10^{-7}\, \rm pc$ scales and driven by a stellar-mass
object fully embedded in the disk and moving subsonically with
respect to the local gas, since we restrict the analysis to circular, equatorial
orbits. Although quasi-periodic eruptions/oscillations in the soft X-ray
have been proposed as signatures of  EMRI-like
system~~\citep{Linial:2023nqs, Franchini:2023bou, Kejriwal:2024bna, Chakraborty:2025xch, Liu:2026tvy, Pelle:2026ohn, Allievi:2026ycd}, these models
typically require the secondary to be at orbital separations of $\gtrsim 100M$, i.e., 
larger than those probed during the final years of inspiral in the
LISA band; and to be on an inclined orbit, so as to
produce a sufficiently strong disk-crossing imprint on the
AGN spectrum~\citep{QPE_plus_GW_detectability}.

Nonetheless, as suggested in~\citet{Duque:2024mfw} for eccentric EMRIs, which also help break the $(\alpha, f_{\rm Edd})$ degeneracy even in a Newtonian framework, the measurement of the massive BH spin and $f_{\rm Edd}$ from GWs alone can be used to estimate the bolometric luminosity of the galaxy. This estimate can then be  cross-matched against AGN catalogs to exclude candidates whose observed luminosity is inconsistent with it \citep{neronov2025catalogveryhighenergyemittingactive, Wise_AGN_catalog, AGN_catalog_2}. This was further explored in~\citet{Lyu:2024gnk, Liu:2026dug} to improve cosmological inference by using EMRIs as dark sirens~\citep{Laghi:2021pqk, laghi2021gravitationalwavecosmologyemris, Liu_2025}. In the absence of a unique electromagnetic
counterpart, the redshift of the source is statistically inferred
by associating the GW sky-localization volume with the host
galaxies present in an AGN catalog, and the luminosity-based
exclusion above reduces the number of candidate hosts. These works estimate that identifying the EMRI as embedded in a disk from GW
observations allows one to reduce the number of potential
host galaxies in the EMRI's sky-localization region by an order of magnitude.

That analysis, however, relied on an assumption of completeness of
the available AGN catalogs, but at present these are highly incomplete, particularly at the redshifts $z \gtrsim 0.3$ relevant for the bulk of LISA EMRI detections. This picture is
expected to change with the Legacy Survey of Space and Time from the Vera C. Rubin Observatory \citep{Vera_Rubin_desc}, which should identify $\sim 10^7$ AGNs up to $z\sim 7$ \citep{Shen_2020, 2021MNRAS.505.5012K}, and characterize AGN variability, providing a large data set that can be cross-checked
against the GW-inferred disk parameters and source locations. Work
is already underway to identify electromagnetic precursors of
LISA massive BH binaries in this same
survey~\citep{Xin:2024fci, Xin:2025voy, Chiesa:2025gwk}.

\section{Discussion}\label{sec:Concl}

In this work, we present the first study of the impact of relativistic corrections to the gas torques on LISA observations of circular EMRIs embedded in an accretion disk. In contrast with forecasts based on Newtonian models~~\citet{Speri:2022upm, Khalvati_2025}, we find that both the typical disk surface density and the accretion rate can be constrained. This suggests that LISA could help distinguish between different gas-flow models in the strong-field gravity regime, on scales much smaller than those probed by typical electromagnetic observations, and could also find applications in multimessenger astronomy, where the inferred disk parameters would help identify the EMRI's host galaxy by cross-correlation with AGN catalogues~\citet{Lyu:2024gnk, Liu:2026dug}. These results highlight both the necessity and the rich scientific return of relativistic modelling of EMRIs in astrophysical environments~\citep{Cardoso:2021wlq, Cardoso:2022whc, Duque:2023seg, Dyson:2025dlj, Li:2025ffh, Vicente:2025gsg, Xu:2026cky, Lui:2026uai}.

Our work also showcases the importance of continued development of fully Bayesian tools for LISA data analysis, as the estimates based on the linear-signal approximation with the Fisher information matrix proved unreliable. This may be due to instabilities in the inversion of ill-conditioned matrices, but the posteriors obtained with MCMC sampling also exhibit non-Gaussian features in the subspace of the highly correlated disk parameters. We leave a more thorough exploration of these aspects to future work, in particular whether higher-order terms in the Fisher matrix formalism could yield better agreement with the MCMC results. In any case, based on the five fiducial EMRI–disk configurations we analysed, we provide an empirical fit that forecasts the constraints on the disk parameters from the dephasing of the signal relative to an EMRI in vacuum.

Nonetheless, this is a first study of its kind and many interesting directions remain to be pursued, with ongoing efforts already underway for some. On the modelling side: (i) include pressure effects self-consistently by solving the relativistic fluid
equations in the EMRI's perturbed spacetime. \cite{Dyson:2026ddd} recently did
this for a restricted disk configuration, but covering the full range of disk
parameters needed for parameter estimation is still an open problem (ii) extend the modelling to high-accretion-rate systems, where the disk puffs up and the thin-disk approximation breaks down~\citep{Slim_acc_disks, 2011arXiv1108.0396S, Kao:2025gxz}; (iii) compare with numerical simulations and include stochastic torque effects (e.g., from turbulence)~\citep{Derdzinski:2020wlw, Derdzinski:2025cql, Garg:2026jrv} within the multiscale-expansion framework for EMRIs; (iv) incorporate coorbital resonances, which become relevant at larger mass ratios~\citep{Derdzinski:2025cql}. On the analysis side, beyond improving individual measurements, the central challenge is extracting beyond-vacuum GR effects, whether from accretion disks, dark matter, or high-energy corrections to GR, in the context of the global fit required to disentangle the $\sim 10^7$ overlapping signals in the LISA data stream~\cite{Littenberg:2023xpl, Katz:2024oqg, Strub:2024kbe, Deng:2025wgk}. This will require: (i) developing model-agnostic waveform models (e.g., by adapting the parametrized post-Einsteinian formalism~\citep{Yunes:2009ke} for EMRIs \citep{Gair_2011} that capture these different physical effects while remaining cheap to evaluate and robust against astrophysical uncertainties; (ii) understanding how to post-process source catalogues obtained with vacuum-only models, by identifying the events of greatest interest for searches of beyond-vacuum GR physics and reusing the posterior samples from the initial inference, for example, by reweighting them under a new model as suggested in~\citet{Kejriwal:2025upp}; (iii) integrating these effects into population synthesis and inference~\cite{Kejriwal:2025jao, Criswell:2026xqk, Toubiana:2026yml}, which is more powerful for detecting small deviations from vacuum GR than individual events. We plan to explore some of these directions in future work.

\begin{appendix}

\appendix

\section{Full corner}\label{sec:App_FullCorner}
\begin{figure*}
     \centering
     \includegraphics[width=\textwidth,height=\textheight,keepaspectratio]{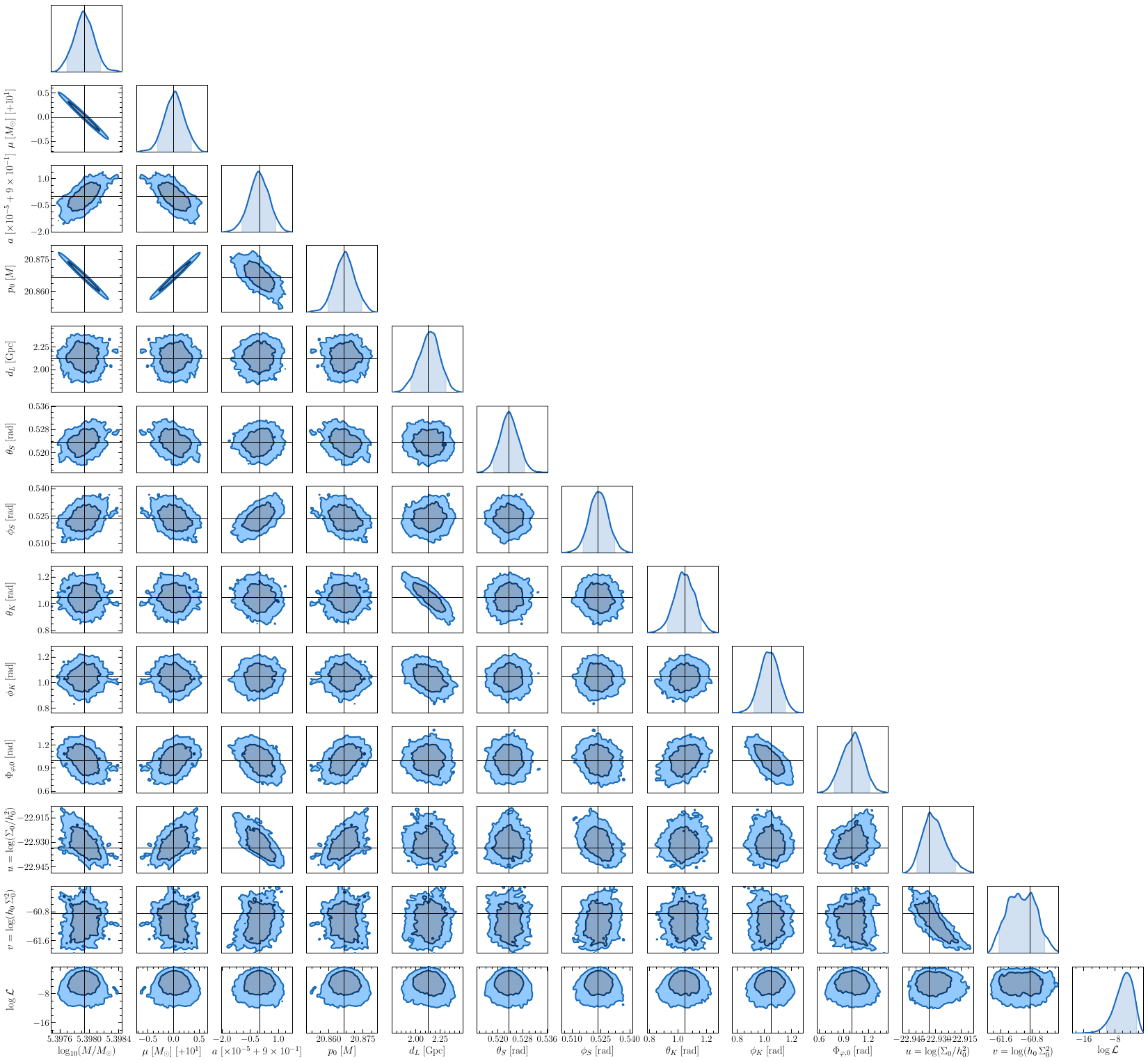}
     \caption{Marginalised 2D posterior distribution across all parameters for the \textit{strong} EMRI-disk system listed in Table~\ref{tab:Systems_summary}. The parameters of the environment are encoded in sampling space through their combinations: $u = log(\Sigma_0/h_0^2)$ and $v = log(h_0 \Sigma_0^2)$.}
     \label{fig:Full_Strong_case}
\end{figure*}

In Fig.~\ref{fig:Full_Strong_case}, we show the complete corner plot for the \textit{strong} configuration of Table~\ref{tab:Systems_summary}.
%
\section{Dephasing across parameter space}\label{sec:App_Dephasing}

\begin{figure*}
\includegraphics[width=0.99\linewidth]{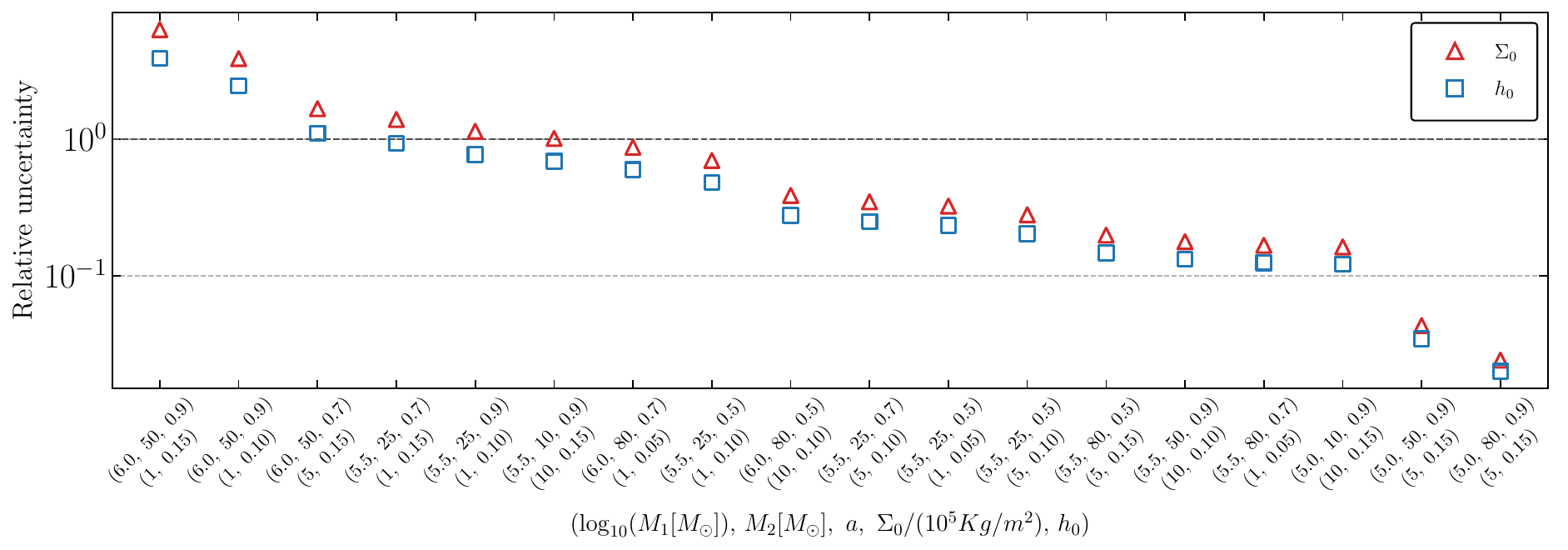}
\caption{Relative uncertainties computed with the power-law expressions introduced in Sec. \ref{sec:results} for several  EMRI-disk configurations}
\label{fig:Many_dephasings}
\end{figure*}

We use $\tt few$ and the modification to compute the dephasing of the waveforms at plunge between vacuum systems and accretion disk systems. Given the dephasing $\Delta \Phi$, we can predict the relative uncertainty adopting the power-law scaling $U(\Delta \Phi)$ for each environmental parameter introduced in Sec. \ref{sec:results}. We illustrate the results for several disk and EMRI configurations in Fig.~\ref{fig:Many_dephasings}. The relative uncertainties are computed for $4$ year long inspirals (coincident with a 4 year observation time). We find that reducing the inspiral time (and observation time) down to 2 years leads to an overall increase of the relative uncertainties by a factor of $\sim 2$.

\end{appendix}

\section*{Acknowledgements}
Part of this work was supported by the German
\emph{Deut\-sche For\-schungs\-ge\-mein\-schaft, DFG\/} project
number Ts~17/2--1. 

\bibliographystyle{mnras}
\bibliography{Refs}

\begin{thebibliography}{}
\makeatletter
\relax
\def\mn@urlcharsother{\let\do\@makeother \do\$\do\&\do\#\do\^\do\_\do\%\do\~}
\def\mn@doi{\begingroup\mn@urlcharsother \@ifnextchar [ {\mn@doi@}
  {\mn@doi@[]}}
\def\mn@doi@[#1]#2{\def\@tempa{#1}\ifx\@tempa\@empty \href
  {http://dx.doi.org/#2} {doi:#2}\else \href {http://dx.doi.org/#2} {#1}\fi
  \endgroup}
\def\mn@eprint#1#2{\mn@eprint@#1:#2::\@nil}
\def\mn@eprint@arXiv#1{\href {http://arxiv.org/abs/#1} {{\tt arXiv:#1}}}
\def\mn@eprint@dblp#1{\href {http://dblp.uni-trier.de/rec/bibtex/#1.xml}
  {dblp:#1}}
\def\mn@eprint@#1:#2:#3:#4\@nil{\def\@tempa {#1}\def\@tempb {#2}\def\@tempc
  {#3}\ifx \@tempc \@empty \let \@tempc \@tempb \let \@tempb \@tempa \fi \ifx
  \@tempb \@empty \def\@tempb {arXiv}\fi \@ifundefined
  {mn@eprint@\@tempb}{\@tempb:\@tempc}{\expandafter \expandafter \csname
  mn@eprint@\@tempb\endcsname \expandafter{\@tempc}}}

\bibitem[\protect\citeauthoryear{Abac et~al.}{Abac
  et~al.}{2025a}]{LIGOScientific:2025pvj}
Abac A.~G.,  et~al., 2025a

\bibitem[\protect\citeauthoryear{Abac et~al.,}{Abac et~al.}{2025b}]{Abac_2025}
Abac A.~G.,  et~al., 2025b, \mn@doi [The Astrophysical Journal Letters]
  {10.3847/2041-8213/ae0c9c}, 993, L25

\bibitem[\protect\citeauthoryear{Abramowicz \& Fragile}{Abramowicz \&
  Fragile}{2013}]{Abramowicz:2011xu}
Abramowicz M.~A.,  Fragile P.~C.,  2013, \mn@doi [Living Rev. Rel.]
  {10.12942/lrr-2013-1}, 16, 1

\bibitem[\protect\citeauthoryear{{Abramowicz}, {Czerny}, {Lasota}  \&
  {Szuszkiewicz}}{{Abramowicz} et~al.}{1988}]{Slim_acc_disks}
{Abramowicz} M.~A.,  {Czerny} B.,  {Lasota} J.~P.,   {Szuszkiewicz} E.,  1988,
  \mn@doi [\apj] {10.1086/166683}, \href
  {https://ui.adsabs.harvard.edu/abs/1988ApJ...332..646A} {332, 646}

\bibitem[\protect\citeauthoryear{Allievi, Broggi, Sesana  \& Bonetti}{Allievi
  et~al.}{2026}]{Allievi:2026ycd}
Allievi C.~M.,  Broggi L.,  Sesana A.,   Bonetti M.,  2026

\bibitem[\protect\citeauthoryear{Antonini, Romero-Shaw, Callister, Dosopoulou,
  Chattopadhyay, Ginat, Gieles  \& Mapelli}{Antonini
  et~al.}{2025}]{Antonini:2025ilj}
Antonini F.,  Romero-Shaw I.,  Callister T.,  Dosopoulou F.,  Chattopadhyay D.,
   Ginat Y.~B.,  Gieles M.,   Mapelli M.,  2025

\bibitem[\protect\citeauthoryear{{Artymowicz}}{{Artymowicz}}{1993a}]{1993ApJ...419..155A}
{Artymowicz} P.,  1993a, \mn@doi [\apj] {10.1086/173469}, \href
  {https://ui.adsabs.harvard.edu/abs/1993ApJ...419..155A} {419, 155}

\bibitem[\protect\citeauthoryear{{Artymowicz}}{{Artymowicz}}{1993b}]{Artymowicz:1993}
{Artymowicz} P.,  1993b, \mn@doi [\apj] {10.1086/173469}, \href
  {https://ui.adsabs.harvard.edu/abs/1993ApJ...419..155A} {419, 155}

\bibitem[\protect\citeauthoryear{{Assef}, {Stern}, {Noirot}, {Jun}, {Cutri}  \&
  {Eisenhardt}}{{Assef} et~al.}{2018}]{Wise_AGN_catalog}
{Assef} R.~J.,  {Stern} D.,  {Noirot} G.,  {Jun} H.~D.,  {Cutri} R.~M.,
  {Eisenhardt} P.~R.~M.,  2018, \mn@doi [\apjs] {10.3847/1538-4365/aaa00a},
  \href {https://ui.adsabs.harvard.edu/abs/2018ApJS..234...23A} {234, 23}

\bibitem[\protect\citeauthoryear{Babak et~al.,}{Babak
  et~al.}{2017}]{Babak_2017_EMRIscience}
Babak S.,  et~al., 2017, \mn@doi [Physical Review D]
  {10.1103/physrevd.95.103012}, 95

\bibitem[\protect\citeauthoryear{Barack \& Pound}{Barack \&
  Pound}{2019}]{Barack:2018yvs}
Barack L.,  Pound A.,  2019, \mn@doi [Rept. Prog. Phys.]
  {10.1088/1361-6633/aae552}, 82, 016904

\bibitem[\protect\citeauthoryear{Barausse, Cardoso  \& Pani}{Barausse
  et~al.}{2014}]{Barausse:2014tra}
Barausse E.,  Cardoso V.,   Pani P.,  2014, \mn@doi [Phys. Rev. D]
  {10.1103/PhysRevD.89.104059}, 89, 104059

\bibitem[\protect\citeauthoryear{{Bardeen} \& {Petterson}}{{Bardeen} \&
  {Petterson}}{1975}]{1975ApJ...195L..65B}
{Bardeen} J.~M.,  {Petterson} J.~A.,  1975, \mn@doi [\apjl] {10.1086/181711},
  \href {https://ui.adsabs.harvard.edu/abs/1975ApJ...195L..65B} {195, L65}

\bibitem[\protect\citeauthoryear{Boumerdassi, Edwards, Vajpeyi  \&
  Burke}{Boumerdassi et~al.}{2025}]{Boumerdassi:2025gvf}
Boumerdassi A.,  Edwards M.~C.,  Vajpeyi A.,   Burke O.,  2025, ]
  {10.1103/9678-764y}

\bibitem[\protect\citeauthoryear{Burke, Marsat, Gair  \& Katz}{Burke
  et~al.}{2025}]{Burke:2025bun}
Burke O.,  Marsat S.,  Gair J.~R.,   Katz M.~L.,  2025, \mn@doi [Phys. Rev. D]
  {10.1103/5jr8-k2ss}, 111, 124053

\bibitem[\protect\citeauthoryear{Burke, Muratore  \& Woan}{Burke
  et~al.}{2026}]{Burke:2025fvl}
Burke O.,  Muratore M.,   Woan G.,  2026, \mn@doi [Phys. Rev. Applied]
  {10.1103/xpvf-syrw}, 25, 034041

\bibitem[\protect\citeauthoryear{Cardoso, Destounis, Duque, Macedo  \&
  Maselli}{Cardoso et~al.}{2022a}]{Cardoso:2021wlq}
Cardoso V.,  Destounis K.,  Duque F.,  Macedo R.~P.,   Maselli A.,  2022a,
  \mn@doi [Phys. Rev. D] {10.1103/PhysRevD.105.L061501}, 105, L061501

\bibitem[\protect\citeauthoryear{Cardoso, Destounis, Duque, Panosso~Macedo  \&
  Maselli}{Cardoso et~al.}{2022b}]{Cardoso:2022whc}
Cardoso V.,  Destounis K.,  Duque F.,  Panosso~Macedo R.,   Maselli A.,  2022b,
  \mn@doi [Phys. Rev. Lett.] {10.1103/PhysRevLett.129.241103}, 129, 241103

\bibitem[\protect\citeauthoryear{Chakraborty et~al.}{Chakraborty
  et~al.}{2025}]{Chakraborty:2025xch}
Chakraborty J.,  et~al., 2025, \mn@doi [Astrophys. J.]
  {10.3847/1538-4357/ae003b}, 992, 120

\bibitem[\protect\citeauthoryear{Chapman-Bird et~al.,}{Chapman-Bird
  et~al.}{2025}]{Chapman_Bird_2025}
Chapman-Bird C.~E.,  et~al., 2025, \mn@doi [Physical Review D]
  {10.1103/scbp-75pf}, 112

\bibitem[\protect\citeauthoryear{Chiesa, Izquierdo-Villalba, Sesana,
  Cocchiararo, Franchini, Lupi, Spinoso  \& Bonoli}{Chiesa
  et~al.}{2026}]{Chiesa:2025gwk}
Chiesa A.,  Izquierdo-Villalba D.,  Sesana A.,  Cocchiararo F.,  Franchini A.,
  Lupi A.,  Spinoso D.,   Bonoli S.,  2026, \mn@doi [Astron. Astrophys.]
  {10.1051/0004-6361/202557029}, 707, A89

\bibitem[\protect\citeauthoryear{Cole, Bertone, Coogan, Gaggero, Karydas,
  Kavanagh, Spieksma  \& Tomaselli}{Cole et~al.}{2023}]{Cole:2022yzw}
Cole P.~S.,  Bertone G.,  Coogan A.,  Gaggero D.,  Karydas T.,  Kavanagh B.~J.,
   Spieksma T. F.~M.,   Tomaselli G.~M.,  2023, \mn@doi [Nature Astron.]
  {10.1038/s41550-023-01990-2}, 7, 943

\bibitem[\protect\citeauthoryear{Cole, Alvey, Speri, Weniger, Bhardwaj, Gerosa
  \& Bertone}{Cole et~al.}{2026}]{Cole:2025sqo}
Cole P.~S.,  Alvey J.,  Speri L.,  Weniger C.,  Bhardwaj U.,  Gerosa D.,
  Bertone G.,  2026, \mn@doi [Phys. Rev. D] {10.1103/4cd3-wfjr}, 113, 063030

\bibitem[\protect\citeauthoryear{Colpi et~al.,}{Colpi
  et~al.}{2024}]{LISA_red_book}
Colpi M.,  et~al., 2024, LISA Definition Study Report (\mn@eprint {arXiv}
  {2402.07571}), \url {https://arxiv.org/abs/2402.07571}

\bibitem[\protect\citeauthoryear{Copparoni, Chandramouli  \&
  Barausse}{Copparoni et~al.}{2025a}]{Copparoni:2025vty}
Copparoni L.,  Chandramouli R.~S.,   Barausse E.,  2025a

\bibitem[\protect\citeauthoryear{Copparoni, Barausse, Speri, Sberna  \&
  Derdzinski}{Copparoni et~al.}{2025b}]{Copparoni:2025jhq}
Copparoni L.,  Barausse E.,  Speri L.,  Sberna L.,   Derdzinski A.,  2025b,
  \mn@doi [Phys. Rev. D] {10.1103/PhysRevD.111.104079}, 111, 104079

\bibitem[\protect\citeauthoryear{Criswell, Banagiri, Delfavero,
  Bustamante-Rosell, Taylor  \& Rosati}{Criswell
  et~al.}{2026}]{Criswell:2026xqk}
Criswell A.~W.,  Banagiri S.,  Delfavero V.,  Bustamante-Rosell M.~J.,  Taylor
  S.~R.,   Rosati R.,  2026

\bibitem[\protect\citeauthoryear{Deng, Babak, Le~Jeune, Marsat, Plagnol  \&
  Sartirana}{Deng et~al.}{2025}]{Deng:2025wgk}
Deng S.,  Babak S.,  Le~Jeune M.,  Marsat S.,  Plagnol {\'E}.,   Sartirana A.,
  2025, \mn@doi [Phys. Rev. D] {10.1103/PhysRevD.111.103014}, 111, 103014

\bibitem[\protect\citeauthoryear{Derdzinski \& Mayer}{Derdzinski \&
  Mayer}{2023}]{Derdzinski:2022ltb}
Derdzinski A.,  Mayer L.,  2023, \mn@doi [Mon. Not. Roy. Astron. Soc.]
  {10.1093/mnras/stad749}, 521, 4522

\bibitem[\protect\citeauthoryear{Derdzinski, D'Orazio, Duffell, Haiman  \&
  MacFadyen}{Derdzinski et~al.}{2021}]{Derdzinski:2020wlw}
Derdzinski A.,  D'Orazio D.,  Duffell P.,  Haiman Z.,   MacFadyen A.,  2021,
  \mn@doi [Mon. Not. Roy. Astron. Soc.] {10.1093/mnras/staa3976}, 501, 3540

\bibitem[\protect\citeauthoryear{Derdzinski et~al.}{Derdzinski
  et~al.}{2025}]{Derdzinski:2025cql}
Derdzinski A.,  et~al., 2025

\bibitem[\protect\citeauthoryear{Duque, Macedo, Vicente  \& Cardoso}{Duque
  et~al.}{2024}]{Duque:2023seg}
Duque F.,  Macedo C. F.~B.,  Vicente R.,   Cardoso V.,  2024, \mn@doi [Phys.
  Rev. Lett.] {10.1103/PhysRevLett.133.121404}, 133, 121404

\bibitem[\protect\citeauthoryear{Duque, Sberna, Spiers  \& Vicente}{Duque
  et~al.}{2025a}]{Duque:2025yfm}
Duque F.,  Sberna L.,  Spiers A.,   Vicente R.,  2025a

\bibitem[\protect\citeauthoryear{Duque, Kejriwal, Sberna, Speri  \& Gair}{Duque
  et~al.}{2025b}]{Duque:2024mfw}
Duque F.,  Kejriwal S.,  Sberna L.,  Speri L.,   Gair J.,  2025b, \mn@doi
  [Phys. Rev. D] {10.1103/PhysRevD.111.084006}, 111, 084006

\bibitem[\protect\citeauthoryear{Dyson \& D'Orazio}{Dyson \&
  D'Orazio}{2026}]{Dyson:2026ddd}
Dyson C.,  D'Orazio D.~J.,  2026

\bibitem[\protect\citeauthoryear{Dyson, Spieksma, Brito, van~de Meent  \&
  Dolan}{Dyson et~al.}{2025}]{Dyson:2025dlj}
Dyson C.,  Spieksma T. F.~M.,  Brito R.,  van~de Meent M.,   Dolan S.,  2025,
  \mn@doi [Phys. Rev. Lett.] {10.1103/PhysRevLett.134.211403}, 134, 211403

\bibitem[\protect\citeauthoryear{{Fairbairn} \& {Dittmann}}{{Fairbairn} \&
  {Dittmann}}{2025}]{2025MNRAS.543..565F}
{Fairbairn} C.~W.,  {Dittmann} A.~J.,  2025, \mn@doi [\mnras]
  {10.1093/mnras/staf1399}, \href
  {https://ui.adsabs.harvard.edu/abs/2025MNRAS.543..565F} {543, 565}

\bibitem[\protect\citeauthoryear{Franchini et~al.,}{Franchini
  et~al.}{2023}]{Franchini:2023bou}
Franchini A.,  et~al., 2023, \mn@doi [Astron. Astrophys.]
  {10.1051/0004-6361/202346565}, 675, A100

\bibitem[\protect\citeauthoryear{Gair \& Yunes}{Gair \&
  Yunes}{2011}]{Gair_2011}
Gair J.,  Yunes N.,  2011, \mn@doi [Physical Review D]
  {10.1103/physrevd.84.064016}, 84

\bibitem[\protect\citeauthoryear{Garg, Derdzinski, Zwick, Capelo  \&
  Mayer}{Garg et~al.}{2022}]{Garg:2022nko}
Garg M.,  Derdzinski A.,  Zwick L.,  Capelo P.~R.,   Mayer L.,  2022, \mn@doi
  [Mon. Not. Roy. Astron. Soc.] {10.1093/mnras/stac2711}, 517, 1339

\bibitem[\protect\citeauthoryear{Garg, Mayer, Wu, Ali-Ha{\"\i}moud  \&
  Lin}{Garg et~al.}{2026}]{Garg:2026jrv}
Garg M.,  Mayer L.,  Wu Y.,  Ali-Ha{\"\i}moud Y.,   Lin D. N.~C.,  2026

\bibitem[\protect\citeauthoryear{{Goldreich} \& {Tremaine}}{{Goldreich} \&
  {Tremaine}}{1980}]{GoldreichTremaine:1980}
{Goldreich} P.,  {Tremaine} S.,  1980, \mn@doi [\apj] {10.1086/158356}, \href
  {https://ui.adsabs.harvard.edu/abs/1980ApJ...241..425G} {241, 425}

\bibitem[\protect\citeauthoryear{{Heckman} \& {Best}}{{Heckman} \&
  {Best}}{2014}]{2014ARA&A..52..589H}
{Heckman} T.~M.,  {Best} P.~N.,  2014, \mn@doi [\araa]
  {10.1146/annurev-astro-081913-035722}, \href
  {https://ui.adsabs.harvard.edu/abs/2014ARA&A..52..589H} {52, 589}

\bibitem[\protect\citeauthoryear{Hegade K.~R., Gammie  \& Yunes}{Hegade K.~R.
  et~al.}{2025a}]{Hegade_model}
Hegade K.~R. A.,  Gammie C.~F.,   Yunes N.,  2025a, \mn@doi [Phys. Rev. D]
  {10.1103/9src-p7sp}, 112, 124012

\bibitem[\protect\citeauthoryear{Hegade K.~R., Gammie  \& Yunes}{Hegade K.~R.
  et~al.}{2025b}]{HegadeKR:2025rpr}
Hegade K.~R. A.,  Gammie C.~F.,   Yunes N.,  2025b, \mn@doi [Phys. Rev. D]
  {10.1103/g83s-jdld}, 112, 124068

\bibitem[\protect\citeauthoryear{Hirata}{Hirata}{2011a}]{Hirata_1}
Hirata C.~M.,  2011a, \mn@doi [Mon. Not. Roy. Astron. Soc.]
  {10.1111/j.1365-2966.2011.18617.x}, 414, 3198

\bibitem[\protect\citeauthoryear{Hirata}{Hirata}{2011b}]{Hirata_2}
Hirata C.~M.,  2011b, \mn@doi [Monthly Notices of the Royal Astronomical
  Society] {10.1111/j.1365-2966.2011.18619.x}, 414, 3212–3230

\bibitem[\protect\citeauthoryear{Hogg \& Foreman-Mackey}{Hogg \&
  Foreman-Mackey}{2018}]{Hogg_2018}
Hogg D.~W.,  Foreman-Mackey D.,  2018, \mn@doi [The Astrophysical Journal
  Supplement Series] {10.3847/1538-4365/aab76e}, 236, 11

\bibitem[\protect\citeauthoryear{Hughes, Warburton, Khanna, Chua  \&
  Katz}{Hughes et~al.}{2021}]{Hughes:2021exa}
Hughes S.~A.,  Warburton N.,  Khanna G.,  Chua A. J.~K.,   Katz M.~L.,  2021,
  \mn@doi [Phys. Rev. D] {10.1103/PhysRevD.103.104014}, 103, 104014

\bibitem[\protect\citeauthoryear{{Ivezi{\'c}} et~al.,}{{Ivezi{\'c}}
  et~al.}{2019}]{Vera_Rubin_desc}
{Ivezi{\'c}} {\v{Z}}.,  et~al., 2019, \mn@doi [\apj]
  {10.3847/1538-4357/ab042c}, \href
  {https://ui.adsabs.harvard.edu/abs/2019ApJ...873..111I} {873, 111}

\bibitem[\protect\citeauthoryear{Jiang, Blaes, Stone  \& Davis}{Jiang
  et~al.}{2019}]{Jiang__2019_AGN_sim_accrate}
Jiang Y.-F.,  Blaes O.,  Stone J.~M.,   Davis S.~W.,  2019, \mn@doi [The
  Astrophysical Journal] {10.3847/1538-4357/ab4a00}, 885, 144

\bibitem[\protect\citeauthoryear{Kao, Capelo, Cenci, Mayer, Lupi  \& Sala}{Kao
  et~al.}{2026}]{Kao:2025gxz}
Kao W.-B.,  Capelo P.~R.,  Cenci E.,  Mayer L.,  Lupi A.,   Sala L.,  2026,
  \mn@doi [Mon. Not. Roy. Astron. Soc.] {10.1093/mnras/stag003}, 546, stag003

\bibitem[\protect\citeauthoryear{Karnesis, Katz, Korsakova, Gair  \&
  Stergioulas}{Karnesis et~al.}{2023}]{Eryn_paper}
Karnesis N.,  Katz M.~L.,  Korsakova N.,  Gair J.~R.,   Stergioulas N.,  2023,
  \mn@doi [Mon. Not. Roy. Astron. Soc.] {10.1093/mnras/stad2939}, 526, 4814

\bibitem[\protect\citeauthoryear{Katz, Chua, Speri, Warburton  \& Hughes}{Katz
  et~al.}{2021}]{FEW_paper}
Katz M.~L.,  Chua A. J.~K.,  Speri L.,  Warburton N.,   Hughes S.~A.,  2021,
  \mn@doi [Phys. Rev. D] {10.1103/PhysRevD.104.064047}, 104, 064047

\bibitem[\protect\citeauthoryear{Katz, Karnesis, Korsakova, Gair  \&
  Stergioulas}{Katz et~al.}{2025}]{Katz:2024oqg}
Katz M.~L.,  Karnesis N.,  Korsakova N.,  Gair J.~R.,   Stergioulas N.,  2025,
  \mn@doi [Phys. Rev. D] {10.1103/PhysRevD.111.024060}, 111, 024060

\bibitem[\protect\citeauthoryear{Kejriwal, Witzany, Zajacek, Pasham  \&
  Chua}{Kejriwal et~al.}{2024}]{Kejriwal:2024bna}
Kejriwal S.,  Witzany V.,  Zajacek M.,  Pasham D.~R.,   Chua A. J.~K.,  2024,
  \mn@doi [Mon. Not. Roy. Astron. Soc.] {10.1093/mnras/stae1599}, 532, 2143

\bibitem[\protect\citeauthoryear{Kejriwal, Duque, Chua  \& Gair}{Kejriwal
  et~al.}{2025}]{Kejriwal:2025upp}
Kejriwal S.,  Duque F.,  Chua A. J.~K.,   Gair J.,  2025, \mn@doi [Phys. Rev.
  D] {10.1103/7f8h-w4yz}, 112, 024005

\bibitem[\protect\citeauthoryear{Kejriwal, Barausse  \& Chua}{Kejriwal
  et~al.}{2026}]{Kejriwal:2025jao}
Kejriwal S.,  Barausse E.,   Chua A. J.~K.,  2026, \mn@doi [Phys. Rev. D]
  {10.1103/rnvp-jbj4}, 113, 064001

\bibitem[\protect\citeauthoryear{Kelly, Vestergaard, Fan, Hopkins, Hernquist
  \& Siemiginowska}{Kelly et~al.}{2010}]{Kelly_2010}
Kelly B.~C.,  Vestergaard M.,  Fan X.,  Hopkins P.,  Hernquist L.,
  Siemiginowska A.,  2010, \mn@doi [The Astrophysical Journal]
  {10.1088/0004-637x/719/2/1315}, 719, 1315–1334

\bibitem[\protect\citeauthoryear{Khalvati, Santini, Duque, Speri, Gair, Yang
  \& Brito}{Khalvati et~al.}{2025}]{Khalvati_2025}
Khalvati H.,  Santini A.,  Duque F.,  Speri L.,  Gair J.,  Yang H.,   Brito R.,
   2025, \mn@doi [Physical Review D] {10.1103/physrevd.111.082010}, 111

\bibitem[\protect\citeauthoryear{{King} \& {Pringle}}{{King} \&
  {Pringle}}{2006}]{Chaotic_accretion}
{King} A.~R.,  {Pringle} J.~E.,  2006, \mn@doi [\mnras]
  {10.1111/j.1745-3933.2006.00249.x}, \href
  {https://ui.adsabs.harvard.edu/abs/2006MNRAS.373L..90K} {373, L90}

\bibitem[\protect\citeauthoryear{{Kley} \& {Nelson}}{{Kley} \&
  {Nelson}}{2012}]{2012ARA&A..50..211K}
{Kley} W.,  {Nelson} R.~P.,  2012, \mn@doi [\araa]
  {10.1146/annurev-astro-081811-125523}, \href
  {https://ui.adsabs.harvard.edu/abs/2012ARA&A..50..211K} {50, 211}

\bibitem[\protect\citeauthoryear{Kocsis, Yunes  \& Loeb}{Kocsis
  et~al.}{2011}]{Kocsis:2011dr}
Kocsis B.,  Yunes N.,   Loeb A.,  2011, \mn@doi [Phys. Rev. D]
  {10.1103/PhysRevD.86.049907}, 84, 024032

\bibitem[\protect\citeauthoryear{Kollmeier et~al.,}{Kollmeier
  et~al.}{2006}]{Kollmeier_2006}
Kollmeier J.~A.,  et~al., 2006, \mn@doi [The Astrophysical Journal]
  {10.1086/505646}, 648, 128

\bibitem[\protect\citeauthoryear{{Kova{\v{c}}evi{\'c}}
  et~al.,}{{Kova{\v{c}}evi{\'c}} et~al.}{2021}]{2021MNRAS.505.5012K}
{Kova{\v{c}}evi{\'c}} A.~B.,  et~al., 2021, \mn@doi [\mnras]
  {10.1093/mnras/stab1595}, \href
  {https://ui.adsabs.harvard.edu/abs/2021MNRAS.505.5012K} {505, 5012}

\bibitem[\protect\citeauthoryear{Laghi}{Laghi}{2021}]{laghi2021gravitationalwavecosmologyemris}
Laghi D.,  2021, Gravitational wave cosmology with EMRIs (\mn@eprint {arXiv}
  {2106.02053}), \url {https://arxiv.org/abs/2106.02053}

\bibitem[\protect\citeauthoryear{Laghi, Tamanini, Del~Pozzo, Sesana, Gair,
  Babak  \& Izquierdo-Villalba}{Laghi et~al.}{2021}]{Laghi:2021pqk}
Laghi D.,  Tamanini N.,  Del~Pozzo W.,  Sesana A.,  Gair J.,  Babak S.,
  Izquierdo-Villalba D.,  2021, \mn@doi [Mon. Not. Roy. Astron. Soc.]
  {10.1093/mnras/stab2741}, 508, 4512

\bibitem[\protect\citeauthoryear{{Lenk}, {Labiano}, {Circosta},
  {Alonso-Herrero}  \& {Wylezalek}}{{Lenk} et~al.}{2026}]{2026A&A...707A.110L}
{Lenk} V.,  {Labiano} A.,  {Circosta} C.,  {Alonso-Herrero} A.,   {Wylezalek}
  D.,  2026, \mn@doi [\aap] {10.1051/0004-6361/202554930}, \href
  {https://ui.adsabs.harvard.edu/abs/2026A&A...707A.110L} {707, A110}

\bibitem[\protect\citeauthoryear{Li, Weller, Bourg, LaHaye, Yunes  \& Yang}{Li
  et~al.}{2025}]{Li:2025ffh}
Li D.,  Weller C.,  Bourg P.,  LaHaye M.,  Yunes N.,   Yang H.,  2025, \mn@doi
  [Phys. Rev. D] {10.1103/7l9s-g21j}, 112, 084057

\bibitem[\protect\citeauthoryear{Linial \& Metzger}{Linial \&
  Metzger}{2023}]{Linial:2023nqs}
Linial I.,  Metzger B.~D.,  2023, \mn@doi [Astrophys. J.]
  {10.3847/1538-4357/acf65b}, 957, 34

\bibitem[\protect\citeauthoryear{Littenberg \& Cornish}{Littenberg \&
  Cornish}{2023}]{Littenberg:2023xpl}
Littenberg T.~B.,  Cornish N.~J.,  2023, \mn@doi [Phys. Rev. D]
  {10.1103/PhysRevD.107.063004}, 107, 063004

\bibitem[\protect\citeauthoryear{{Liu}, {Liu}, {Dong}, {Zhou}, {Wang}, {Lu}  \&
  {Yuan}}{{Liu} et~al.}{2019}]{AGN_catalog_2}
{Liu} H.-Y.,  {Liu} W.-J.,  {Dong} X.-B.,  {Zhou} H.,  {Wang} T.,  {Lu} H.,
  {Yuan} W.,  2019, \mn@doi [\apjs] {10.3847/1538-4365/ab298b}, \href
  {https://ui.adsabs.harvard.edu/abs/2019ApJS..243...21L} {243, 21}

\bibitem[\protect\citeauthoryear{Liu, Han  \& Yun}{Liu et~al.}{2025}]{Liu_2025}
Liu J.-D.,  Han W.-B.,   Yun Q.,  2025, \mn@doi [The Astrophysical Journal]
  {10.3847/1538-4357/adfc3e}, 991, 223

\bibitem[\protect\citeauthoryear{Liu, Han  \& Tagawa}{Liu
  et~al.}{2026b}]{Liu:2026dug}
Liu J.-D.,  Han W.-B.,   Tagawa H.,  2026b

\bibitem[\protect\citeauthoryear{Liu, Liu, Pan, Deng, Shen  \& Yu}{Liu
  et~al.}{2026a}]{Liu:2026tvy}
Liu K.,  Liu S.-F.,  Pan Z.,  Deng H.,  Shen R.,   Yu C.,  2026a

\bibitem[\protect\citeauthoryear{{Lops}, {Izquierdo-Villalba}, {Colpi},
  {Bonoli}, {Sesana}  \& {Mangiagli}}{{Lops}
  et~al.}{2023}]{2023MNRAS.519.5962L}
{Lops} G.,  {Izquierdo-Villalba} D.,  {Colpi} M.,  {Bonoli} S.,  {Sesana} A.,
  {Mangiagli} A.,  2023, \mn@doi [\mnras] {10.1093/mnras/stad058}, \href
  {https://ui.adsabs.harvard.edu/abs/2023MNRAS.519.5962L} {519, 5962}

\bibitem[\protect\citeauthoryear{Lui, Drummond  \& Torres-Orjuela}{Lui
  et~al.}{2026}]{Lui:2026uai}
Lui L.,  Drummond L.~V.,   Torres-Orjuela A.,  2026

\bibitem[\protect\citeauthoryear{Lyu, Pan, Mao, Jiang  \& Yang}{Lyu
  et~al.}{2026}]{Lyu:2024gnk}
Lyu Z.,  Pan Z.,  Mao J.,  Jiang N.,   Yang H.,  2026, \mn@doi [Phys. Rev. D]
  {10.1103/337c-g6x1}, 113, 043002

\bibitem[\protect\citeauthoryear{Masset}{Masset}{2002}]{Masset_2002}
Masset F.~S.,  2002, \mn@doi [Astronomy &amp; Astrophysics]
  {10.1051/0004-6361:20020240}, 387, 605–623

\bibitem[\protect\citeauthoryear{Morton, Rinaldi, Torres-Orjuela, Derdzinski,
  Vaccaro  \& Pozzo}{Morton et~al.}{2023}]{morton2023gw190521binaryblackhole}
Morton S.,  Rinaldi S.,  Torres-Orjuela A.,  Derdzinski A.,  Vaccaro M.~P.,
  Pozzo W.~D.,  2023, GW190521: a binary black hole merger inside an active
  galactic nucleus? (\mn@eprint {arXiv} {2310.16025}), \url
  {https://arxiv.org/abs/2310.16025}

\bibitem[\protect\citeauthoryear{{M{\"u}ller}, {Kley}  \& {Meru}}{{M{\"u}ller}
  et~al.}{2012}]{Thin_disk_sim}
{M{\"u}ller} T.~W.~A.,  {Kley} W.,   {Meru} F.,  2012, \mn@doi [\aap]
  {10.1051/0004-6361/201118737}, \href
  {https://ui.adsabs.harvard.edu/abs/2012A&A...541A.123M} {541, A123}

\bibitem[\protect\citeauthoryear{Neronov \& Semikoz}{Neronov \&
  Semikoz}{2025}]{neronov2025catalogveryhighenergyemittingactive}
Neronov A.,  Semikoz D.,  2025, Catalog of very-high-energy emitting active
  galactic nuclei at high Galactic latitudes (\mn@eprint {arXiv} {2506.08497}),
  \url {https://arxiv.org/abs/2506.08497}

\bibitem[\protect\citeauthoryear{{Novikov} \& {Thorne}}{{Novikov} \&
  {Thorne}}{1973}]{1973blho.conf..343N}
{Novikov} I.~D.,  {Thorne} K.~S.,  1973, in {Dewitt} C.,  {Dewitt} B.~S.,  eds,
  Black Holes (Les Astres Occlus). pp 343--450

\bibitem[\protect\citeauthoryear{Pan \& Yang}{Pan \&
  Yang}{2021}]{Pan_2021_EMRI_in_AGN_formation}
Pan Z.,  Yang H.,  2021, \mn@doi [Physical Review D]
  {10.1103/physrevd.103.103018}, 103

\bibitem[\protect\citeauthoryear{Pelle, Kawaguchi, Shibata  \& Lam}{Pelle
  et~al.}{2026}]{Pelle:2026ohn}
Pelle J.,  Kawaguchi K.,  Shibata M.,   Lam A. T.-L.,  2026

\bibitem[\protect\citeauthoryear{{Planck Collaboration}, Aghanim
  et~al.}{{Planck Collaboration} et~al.}{2020}]{Planck:2018vyg}
{Planck Collaboration} Aghanim N.,   et~al., 2020, \mn@doi [Astron. Astrophys.]
  {10.1051/0004-6361/201833910}, 641, A6

\bibitem[\protect\citeauthoryear{Potter}{Potter}{2021}]{Potter_2021}
Potter W.~J.,  2021, \mn@doi [Monthly Notices of the Royal Astronomical
  Society] {10.1093/mnras/stab636}, 503, 5025–5045

\bibitem[\protect\citeauthoryear{Pound \& Wardell}{Pound \&
  Wardell}{2021}]{Pound:2021qin}
Pound A.,  Wardell B.,  2021, ] {10.1007/978-981-15-4702-7\_38-1}

\bibitem[\protect\citeauthoryear{{Sadowski}}{{Sadowski}}{2011}]{2011arXiv1108.0396S}
{Sadowski} A.,  2011, \mn@doi [arXiv e-prints] {10.48550/arXiv.1108.0396},
  \href {https://ui.adsabs.harvard.edu/abs/2011arXiv1108.0396S} {p.
  arXiv:1108.0396}

\bibitem[\protect\citeauthoryear{{Shakura} \& {Sunyaev}}{{Shakura} \&
  {Sunyaev}}{1973}]{1973A&A....24..337S}
{Shakura} N.~I.,  {Sunyaev} R.~A.,  1973, \aap, \href
  {https://ui.adsabs.harvard.edu/abs/1973A&A....24..337S} {24, 337}

\bibitem[\protect\citeauthoryear{Shen, Hopkins, Faucher-Giguère, Alexander,
  Richards, Ross  \& Hickox}{Shen et~al.}{2020}]{Shen_2020}
Shen X.,  Hopkins P.~F.,  Faucher-Giguère C.-A.,  Alexander D.~M.,  Richards
  G.~T.,  Ross N.~P.,   Hickox R.~C.,  2020, \mn@doi [Monthly Notices of the
  Royal Astronomical Society] {10.1093/mnras/staa1381}, 495, 3252–3275

\bibitem[\protect\citeauthoryear{Speri, Antonelli, Sberna, Babak, Barausse,
  Gair  \& Katz}{Speri et~al.}{2023}]{Speri:2022upm}
Speri L.,  Antonelli A.,  Sberna L.,  Babak S.,  Barausse E.,  Gair J.~R.,
  Katz M.~L.,  2023, \mn@doi [Phys. Rev. X] {10.1103/PhysRevX.13.021035}, 13,
  021035

\bibitem[\protect\citeauthoryear{Speri, Duque, Barsanti, Santini, Kejriwal,
  Burke  \& Chapman-Bird}{Speri et~al.}{2026a}]{Speri:2026ade}
Speri L.,  Duque F.,  Barsanti S.,  Santini A.,  Kejriwal S.,  Burke O.,
  Chapman-Bird C. E.~A.,  2026a

\bibitem[\protect\citeauthoryear{Speri, Tenorio, Chapman-Bird  \& Gerosa}{Speri
  et~al.}{2026b}]{Speri:2025ucn}
Speri L.,  Tenorio R.,  Chapman-Bird C.,   Gerosa D.,  2026b, \mn@doi [Phys.
  Rev. D] {10.1103/dh3j-ksfl}, 113, 024061

\bibitem[\protect\citeauthoryear{Spieksma \& Cannizzaro}{Spieksma \&
  Cannizzaro}{2026}]{Spieksma_2026}
Spieksma T. F.~M.,  Cannizzaro E.,  2026, \mn@doi [Monthly Notices of the Royal
  Astronomical Society] {10.1093/mnras/stag021}, 546

\bibitem[\protect\citeauthoryear{Strub, Ferraioli, Schmelzbach, St{\"a}hler  \&
  Giardini}{Strub et~al.}{2024}]{Strub:2024kbe}
Strub S.~H.,  Ferraioli L.,  Schmelzbach C.,  St{\"a}hler S.~C.,   Giardini D.,
   2024, \mn@doi [Phys. Rev. D] {10.1103/PhysRevD.110.024005}, 110, 024005

\bibitem[\protect\citeauthoryear{Strub, Speri  \& Giardini}{Strub
  et~al.}{2026}]{Strub:2025dfs}
Strub S.~H.,  Speri L.,   Giardini D.,  2026, \mn@doi [Phys. Rev. D]
  {10.1103/yp6d-8vqv}, 113, 084048

\bibitem[\protect\citeauthoryear{Suzuguchi, Omiya  \& Takeda}{Suzuguchi
  et~al.}{2025}]{QPE_plus_GW_detectability}
Suzuguchi T.,  Omiya H.,   Takeda H.,  2025, Possibility of Multi-Messenger
  Observations of Quasi-Periodic Eruptions with X-rays and Gravitational Waves
  (\mn@eprint {arXiv} {2505.10488}), \url {https://arxiv.org/abs/2505.10488}

\bibitem[\protect\citeauthoryear{Tanaka \& Okada}{Tanaka \&
  Okada}{2024}]{Tanaka_2024}
Tanaka H.,  Okada K.,  2024, \mn@doi [The Astrophysical Journal]
  {10.3847/1538-4357/ad410d}, 968, 28

\bibitem[\protect\citeauthoryear{{Tanaka}, {Takeuchi}  \& {Ward}}{{Tanaka}
  et~al.}{2002}]{2002ApJ...565.1257T}
{Tanaka} H.,  {Takeuchi} T.,   {Ward} W.~R.,  2002, \mn@doi [\apj]
  {10.1086/324713}, \href
  {https://ui.adsabs.harvard.edu/abs/2002ApJ...565.1257T} {565, 1257}

\bibitem[\protect\citeauthoryear{{Teukolsky}}{{Teukolsky}}{1973}]{1973ApJ...185..635T}
{Teukolsky} S.~A.,  1973, \mn@doi [\apj] {10.1086/152444}, \href
  {https://ui.adsabs.harvard.edu/abs/1973ApJ...185..635T} {185, 635}

\bibitem[\protect\citeauthoryear{{Thorne}}{{Thorne}}{1974}]{Thorne:1974}
{Thorne} K.~S.,  1974, \mn@doi [\apj] {10.1086/152991}, \href
  {https://ui.adsabs.harvard.edu/abs/1974ApJ...191..507T} {191, 507}

\bibitem[\protect\citeauthoryear{Tong et~al.}{Tong et~al.}{2026}]{Tong:2025wpz}
Tong H.,  et~al., 2026, \mn@doi [Nature] {10.1038/s41586-026-10359-0}, 652, 874

\bibitem[\protect\citeauthoryear{Toubiana \& Gair}{Toubiana \&
  Gair}{2026}]{Toubiana:2026yml}
Toubiana A.,  Gair J.,  2026

\bibitem[\protect\citeauthoryear{Toubiana et~al.}{Toubiana
  et~al.}{2021}]{Toubiana:2020drf}
Toubiana A.,  et~al., 2021, \mn@doi [Phys. Rev. Lett.]
  {10.1103/PhysRevLett.126.101105}, 126, 101105

\bibitem[\protect\citeauthoryear{Vicente, Karydas  \& Bertone}{Vicente
  et~al.}{2025}]{Vicente:2025gsg}
Vicente R.,  Karydas T.~K.,   Bertone G.,  2025, \mn@doi [Phys. Rev. Lett.]
  {10.1103/s4wh-x6c4}, 135, 211401

\bibitem[\protect\citeauthoryear{Ward}{Ward}{1997}]{Ward:1997}
Ward W.~R.,  1997, \mn@doi [Icarus] {https://doi.org/10.1006/icar.1996.5647},
  126, 261

\bibitem[\protect\citeauthoryear{Wu, Chen  \& Lin}{Wu
  et~al.}{2023}]{Wu:2023qeh}
Wu Y.,  Chen Y.-X.,   Lin D. N.~C.,  2023, \mn@doi [Mon. Not. Roy. Astron.
  Soc.] {10.1093/mnrasl/slad183}, 528, L127

\bibitem[\protect\citeauthoryear{Xin \& Haiman}{Xin \&
  Haiman}{2024}]{Xin:2024fci}
Xin C.,  Haiman Z.,  2024, \mn@doi [Mon. Not. Roy. Astron. Soc.]
  {10.1093/mnras/stae2009}, 533, 3164

\bibitem[\protect\citeauthoryear{Xin, Isi, Farr  \& Haiman}{Xin
  et~al.}{2026}]{Xin:2025voy}
Xin C.,  Isi M.,  Farr W.~M.,   Haiman Z.,  2026, \mn@doi [Astrophys. J.]
  {10.3847/1538-4357/ae40b3}, 999, 149

\bibitem[\protect\citeauthoryear{Xu, Brito, Della~Monica, Vicente  \& Yuan}{Xu
  et~al.}{2026}]{Xu:2026cky}
Xu Q.-X.,  Brito R.,  Della~Monica R.,  Vicente R.,   Yuan C.,  2026

\bibitem[\protect\citeauthoryear{Yunes \& Pretorius}{Yunes \&
  Pretorius}{2009}]{Yunes:2009ke}
Yunes N.,  Pretorius F.,  2009, \mn@doi [Phys. Rev. D]
  {10.1103/PhysRevD.80.122003}, 80, 122003

\bibitem[\protect\citeauthoryear{Yunes, Kocsis, Loeb  \& Haiman}{Yunes
  et~al.}{2011}]{Yunes:2011ws}
Yunes N.,  Kocsis B.,  Loeb A.,   Haiman Z.,  2011, \mn@doi [Phys. Rev. Lett.]
  {10.1103/PhysRevLett.107.171103}, 107, 171103

\bibitem[\protect\citeauthoryear{Zeng \& Pan}{Zeng \& Pan}{2026}]{Zeng:2026ydj}
Zeng Y.,  Pan Z.,  2026

\bibitem[\protect\citeauthoryear{Zwick, Derdzinski, Garg, Capelo  \&
  Mayer}{Zwick et~al.}{2022}]{Zwick:2021dlg}
Zwick L.,  Derdzinski A.,  Garg M.,  Capelo P.~R.,   Mayer L.,  2022, \mn@doi
  [Mon. Not. Roy. Astron. Soc.] {10.1093/mnras/stac299}, 511, 6143

\makeatother
\end{thebibliography}
\end{document}